Title:    Simulation of *Drosophila* Circadian Oscillations, Mutations, and Light
Responses by a Model with VRI, PDP-1, and CLK

Authors:  Paul Smolen, Paul E. Hardin[*], Brian S. Lo, Douglas A. Baxter, and John H. Byrne

31 pages, 7 figures, 1 table

Laboratories
of Origin:                        Department of Neurobiology and Anatomy
                 W.M. Keck Center for the Neurobiology of Learning and Memory
                      The University of Texas-Houston Medical School
                                    P.O. Box 20708
                                   Houston, TX 77225

                            * Department of Biology and Biochemistry
                               369 Science and Research 2 Bldg.
                                    University of Houston
                                    Houston, TX 77204

Running Title:                    *Drosophila* circadian model

Correspondence Address:                 John H. Byrne
                        Department of Neurobiology and Anatomy
                 W.M. Keck Center for the Neurobiology of Learning and Memory
                      The University of Texas-Houston Medical School
                                    P.O. Box 20708
                                   Houston, TX 77225
                                 Voice: (713) 500-5602
                                  FAX: (713) 500-0623
                             E-mail: John.H.Byrne@uth.tmc.edu

Acknowledgements:

           Supported by NIH grant P01 NS38310 and DARPA grant N00014-01-1-1031.

## Abstract

A model of *Drosophila* circadian rhythm generation was developed to represent feedback loops based on transcriptional regulation of *per, Clk (dclock), Pdp-1,* and *vri (vrille).* The model postulates that histone acetylation kinetics make transcriptional activation a nonlinear function of [CLK]. Such a nonlinearity is essential to simulate robust circadian oscillations of transcription in our model and in previous models. Simulations suggest two positive feedback loops involving *Clk* are not essential for oscillations, because oscillations of [PER] were preserved when *Clk, vri,* or *Pdp-1* expression was fixed. However, eliminating positive feedback by fixing *vri* expression altered the oscillation period. Eliminating the negative feedback loop in which PER represses *per* expression abolished oscillations. Simulations of *per* or *Clk* null mutations, of *per* overexpression, and of *vri, Clk,* or *Pdp-1* heterozygous null mutations altered model behavior in ways similar to experimental data. The model simulated a photic phase-response curve resembling experimental curves, and oscillations entrained to simulated light-dark cycles. Temperature compensation of oscillation period could be simulated if temperature elevation slowed PER nuclear entry or PER phosphorylation. The model makes experimental predictions, some of which could be tested in transgenic *Drosophila.*

## Introduction

Circadian rhythms in physiology and behavior depend on the oscillating expression of genes, a few of which act as core clock components. The first core clock components identified in *Drosophila* were *per* and *tim. Per* and *tim* are activated by a heterodimer of the transcription factors CLK and CYC (Bae et al., 2000; Lee et al., 1999; Allada et al., 1998; Darlington et al., 1998). PER and TIM repress *per* and *tim* transcription, forming a negative feedback loop. This repression arises from binding of PER and TIM or of PER alone to the CLK-CYC heterodimer, preventing activation of *per* and *tim* (Bae et al., 2000; Rothenfluh et al., 2000; Lee et al., 1999). Positive feedback is also present. Elevated CLK represses *Clk* (Glossop et al., 1999). PER indirectly activates *Clk* by binding CLK and blocking this repression (Glossop et al., 1999; Bae et al., 1998). CLK's activation of *per* then closes a positive feedback loop. Another core clock component, *vri* (or *vrille*), is activated by CLK-CYC (Blau and Young, 1999) and in turn represses *Clk* (Glossop et al., 2003). *Par Domain Protein 1 (Pdp-1)* is another core clock component which activates *Clk* and is activated by CLK-CYC (Cyran et al., 2003).

Modeling is important for gaining understanding of the dynamics of gene systems containing complex regulatory motifs, such as multiple feedback loops (Hasty et al., 2001; Smolen et al., 2000). Previous models have considered the negative feedback loop based on *per* and *tim* repression (Leloup et al., 1998; Goldbeter, 1995) and a few models have considered regulation of *Clk* as well as *Clk's* activation of *per* (Smolen et al., 2002; Smolen et al., 2001; Ueda et al., 2001). However, no model has yet considered regulation of *vri* expression, VRI's repression of *Clk*, and the role of *Pdp-1*. In order to encapsulate current understanding and to make predictions that can guide further experiments, we believe it is timely to develop a model that represents the mechanisms of circadian transcriptional regulation as currently understood in *Drosophila*.

Figure 1A schematizes our model's representation of the transcriptional feedback loops. On the left is the *per* negative feedback loop, in which CLK activates *per* expression and PER then represses *Clk*. The TIM gene product is not represented in the model. The rationale for this simplification is discussed below (Model Development and Numerical Methods). On the right is the *vri* negative feedback loop, in which CLK activates *vri* and VRI then represses *Clk*. The positive feedback loop discussed above, involving *Clk* repression, subsumes both these negative feedback loops as follows. When *Clk* is activated, CLK levels increase. Activation of *per* by CLK results in PER synthesis and the binding of CLK by PER. Thus, activation of *vri* by CLK is diminished, and VRI levels fall. Finally, repression of *Clk* by *vri* is relieved, and CLK increases further. On the top of Fig. 1A, the reciprocal activation of *Clk* by the *Pdp-1* gene product and *vice versa* forms a second positive feedback loop. Figure 1A emphasizes the ways in which all the feedback loops include regulation of *Clk* expression or CLK function.

Figure 1B illustrates in more detail the dynamics of PER protein. After synthesis, PER undergoes multiple phosphorylations over a period of hrs (Edery et al., 1994). PER is also transported into the nucleus, where it interacts with CLK, inhibiting transcriptional activation by CLK. Following multiple phosphorylations, PER is degraded, and CLK induction of *per, vri,* and *Pdp-1* again takes place.

Etchegaray et al. (2003) reported that mammalian *per* genes exhibit circadian rhythms in acetylation of histone proteins, synchronous with rhythms of *per* mRNA. CRY proteins inhibit CLK-activated transcription, suggesting CRY inhibits histone acetylation and transcriptional activation by interacting with CLK. Our model assumes *Drosophila* CLK similarly induces histone acetylation, activating transcription of *per, vri,* and *Pdp-1*. PER, rather than CRY, is assumed to inhibit acetylation and repress transcription. Previous models (Ueda et al., 2001; Leloup et al., 1998; Goldbeter, 1995) have found it essential to describe activation of *per* expression with nonlinear Hill functions of [CLK] or [PER]. Hill coefficients of 3-5 were used to create steep, sigmoidal functions of transcription rate *vs.* [CLK] or [PER]. Requirements for CLK or PER to bind to multiple E-box enhancer elements in the *per* promoter region could justify nonlinear Hill functions. However, a promoter fragment containing a single E-box, coupled to a *per* transgene, is sufficient to drive robust *per* mRNA cycling and rescue behavioral rhythmicity in *per[01]* mutants (Hao et al., 1999). Therefore, we suggest kinetics of multiple histone acetylation events are more likely to generate the steep, sigmoidal relationships of *per* expression rate *vs.* levels of CLK and PER that modeling suggests is necessary for robust circadian oscillations. We developed equations to describe these sigmoidal relationships. In the resulting model, as in previous models (Ueda et al., 2001; Leloup et al., 1998; Goldbeter, 1995), we find that if sigmoidicity is removed, circadian oscillations can no longer be simulated.

Our model simulates oscillations in constant darkness, photic entrainment of oscillations, the photic phase-response curve, and temperature compensation of oscillation period. The model also simulates the effects of null mutations of *per* and *Clk*, and the effects of *vri, Clk,* and *Pdp-1* heterozygous null mutations on oscillation period. Simulations suggest the negative feedback loop in which PER interacts with CLK to inhibit *per* expression is essential for circadian oscillations of gene expression.

Simulations also suggest the positive feedback loop involving *Clk* repression is not essential for oscillations in the expression of *per, vri,* or *Pdp-1*. However, this feedback loop is likely to play an essential role in driving oscillations in *Clk* expression and in regulating many clock-controlled genes regulated by CLK (McDonald and Rosbash, 2001). Therefore, this feedback loop is likely to mediate behavioral aspects of rhythmicity. Simulations also suggest the second positive feedback loop of reciprocal *Pdp-1* and *Clk* activation may not be essential for circadian oscillations of the expression of genes other than *Pdp-1*.

The model makes experimental predictions. Expression of a reporter gene with a promoter including CLK binding sites is predicted to depend steeply on levels of PER and/or CLK. Circadian oscillations of *per* expression are predicted to be preserved with constitutive *Pdp-1* or *Clk* expression. Either the rate of PER nuclear entry or that of PER phosphorylation is predicted to <u>not</u> increase with temperature.

## Model Development and Numerical Methods

For simplicity, all concentrations are referenced to the total cell volume. Absolute concentrations of circadian proteins in *Drosophila* neurons have not been determined, although relative abundances of PER, CLK, CYC, and TIM have been quantified (Bae et al., 2000). We chose parameters such that the average concentration of PER during an oscillation has a plausible value, approximately 2 nM. In a *Drosophila* lateral neuron with a radius of 5-6 μm (Ewer et al., 1992), 2 nM would correspond to roughly 1,000 PER molecules.

Circadian expression of *per, vri,* and *Pdp-1* is activated by binding of CLK-CYC heterodimers to E-box sequences (CACGTG) in their promoter regions (Cyran et al., 2003; Blau and Young, 1999; Hao et al., 1999). The concentration of CYC in head extract is relatively constant and more than two orders of magnitude above the concentration of CLK, and most CLK in head extract interacts stably with CYC (Bae et al., 2000). Therefore, it is reasonable to assume that the level of CLK-CYC is limited by [CLK], and to model transcriptional activation as a function of [CLK].

Etchegaray et al. (2003) provide evidence that mammalian CLK regulates transcription of circadian genes by promoting histone acetylation. CLK/BMAL1 heterodimers activate the expression of three *per* genes (*Per1-3*) and two cryptochrome genes (*Cry1-2*). CRY proteins inhibit CLK-activated transcription. The promoter regions of *Per1, Per2* and *Cry1* exhibit circadian rhythms in acetylation of H3 histone proteins and RNA polymerase II binding. These rhythms are synchronous with those of *Per1, Per2* and *Cry1* mRNA. The histone acetyltransferase p300 precipitates with CLK in a circadian manner. CRY proteins inhibit the p300-induced increase in CLK-activated transcription. These results suggest CLK interacts with p300 to promote histone acetylation, chromatin structure alteration, and transcriptional activation, whereas CRY interacts with CLK-p300 to inhibit acetylation and reduce transcription. It is plausible that regulation of circadian transcription by *Drosophila* CLK and PER also involves histone acetylation by p300 histone acetyltransferase. In *Drosophila*, CRY proteins do not appear important for repressing CLK-activated transcription. Instead, the interaction of PER with CLK may reduce histone acetyltransferase activity, allowing histone deacetylation. We developed a phenomenological representation of multiple histone acetylations that generates steep, sigmoidal relationships of *per*, *vri*, and *Pdp-1* expression rates *vs.* [CLK] and [PER]. As discussed in the Introduction, previous models have suggested such nonlinear relationships are essential for robust circadian oscillations.

We introduce first-order rate constants to describe acetylation of amino acid residues in histone proteins near E-box elements in the *vri*, *per*, and *Pdp-1* promoters. These rate constants are respectively denoted $k_{Vacet}$, $k_{Pacet}$, and $k_{PDacet}$. Acetylation is assumed to require the binding of CLK to an E-box, and to be repressed by binding of PER. The acetylation rate constants are therefore represented as a product of hyperbolic Hill functions of [CLK] and of [PER]. The latter function describes repression by nuclear PER with concentration [$PER_{nuc}$]. The following expressions result,

$$k_{Vacet} = F_V \frac{[CLK]}{[CLK] + K_{CV}} \frac{K_{PV}}{[PER_{nuc}] + K_{PV}} \qquad 1)$$

$$k_{Pacet} = F_P \frac{[CLK]}{[CLK] + K_{CP}} \frac{K_{PP}}{[PER_{nuc}] + K_{PP}} \qquad 2)$$

$$k_{PDacet} = F_{PD} \frac{[CLK]}{[CLK] + K_{CPD}} \frac{K_{PPD}}{[PER_{nuc}] + K_{PPD}} \qquad 3)$$

In Eqs. 1-3, $F_V$, $F_P$, and $F_{PD}$ are arbitrary scaling factors. Corresponding rate constants for deacetylation at the *vri, per,* and *Pdp-1* promoters are introduced. These are denoted $k_{Vdeac}$, $k_{Pdeac}$, and $k_{PDdeac}$, and are assumed to be fixed. This assumption is consistent with data of Etchegaray et al. (2003) illustrating non-rhythmic histone deacetylase activity at the mammalian *per1* and *per2* promoters.

Variables $AC_{Vri}$, $AC_{Per}$, and $AC_{Pdp}$ describe acetylation of amino acid residues in a representative histone protein at the *vri, per,* or *Pdp-1* promoters. These variables denote the fraction of eligible residues that are acetylated. Other rhythmic histone modifications, such as phosphorylations (Etchegaray et al., 2003), may also affect transcription and could be subsumed into $AC_{Vri}$, $AC_{Per}$, and $AC_{Pdp}$. Because the number of residues eligible for acetylation or for other rhythmic modification is not known, we approximate $AC_{Vri}$, $AC_{Per}$, and $AC_{Pdp}$ as continuous variables ranging from 0 to 1. First-order differential equations describe their dynamics, using acetylation rate constants defined by Eqs. 1-3 and the corresponding deacetylation rate constants,

$$\frac{d(AC_{Vri})}{dt} = k_{Vacet}(1 - AC_{Vri}) - k_{Vdeac}AC_{Vri} \qquad 4)$$

$$\frac{d(AC_{Per})}{dt} = k_{Pacet}(1 - AC_{Per}) - k_{Pdeac}AC_{Per} \qquad 5)$$

$$\frac{d(AC_{Pdp})}{dt} = k_{PDacet}(1 - AC_{Pdp}) - k_{PDdeac}AC_{Pdp} \qquad 6)$$

These equations do not yet imply any sigmoidal relationships between levels of CLK or PER and transcriptional activation. To introduce sigmoidicity, we make the plausible assumption that multiple histone proteins need to be acetylated at the *vri, per,* and *Pdp-1* promoters in order to bring about an "open" DNA configuration readily accessible to RNA polymerase. We describe accessibility to RNA polymerase by variables $OP_{Vri}$, $OP_{Per}$, and $OP_{Pdp}$. A requirement for acetylation of multiple histones can be represented as a nonlinear dependence of $OP_{Vri}$, $OP_{Per}$, and $OP_{Pdp}$ on the single-histone acetylation fractions $AC_{Vri}$, $AC_{Per}$, and $AC_{Pdp}$. We chose a phenomenological, sigmoidal nonlinearity. Accessibilities were assumed to approach steady-state values given by a power N of the acetylation fractions (usually N=5). These steady-state values were denoted $OP_{Vri, ss}$, $OP_{Per, ss}$, and $OP_{Pdp, ss}$,

$$OP_{Vri, ss} = \left(AC_{Vri}\right)^{N} \qquad\qquad 7)$$

$$OP_{Per, ss} = \left(AC_{Per}\right)^{N} \qquad\qquad 8)$$

$$OP_{Pdp, ss} = \left(AC_{Pdp}\right)^{N} \qquad\qquad 9)$$

$OP_{Vri}$, $OP_{Per}$, and $OP_{Pdp}$ relax to these steady-state values with time constants $\tau_{Vri, op}$, $\tau_{Per, op}$, and $\tau_{Pdp, op}$,

$$\frac{d\left(OP_{Vri}\right)}{dt} = \frac{OP_{Vri} - OP_{Vri, ss}}{\tau_{Vri, op}} \qquad\qquad 10)$$

$$\frac{d\left(OP_{Per}\right)}{dt} = \frac{OP_{Per} - OP_{Per, ss}}{\tau_{Per, op}} \qquad\qquad 11)$$

$$\frac{d\left(OP_{Pdp}\right)}{dt} = \frac{OP_{Pdp} - OP_{Pdp, ss}}{\tau_{Pdp, op}} \qquad\qquad 12)$$

The expression rates of *per, vri,* and *Pdp-1* are denoted $R_{Per}$, $R_{Vri}$, and $R_{Pdp}$. These rates are assumed proportional to the accessibility of promoters to RNA polymerase, *i.e.*, to $OP_{Per}$, $OP_{Vri}$, and $OP_{Pdp}$. Small basal rates in the absence of CLK are denoted $R_{Pbas}$, $R_{Vbas}$, and $R_{PDbas}$. The following equations result

$$R_{Per} = V_{Per}OP_{Per} + R_{Pbas} \qquad\qquad 13)$$

$$R_{Vri} = V_{Vri}OP_{Vri} + R_{Vbas} \qquad\qquad 14)$$

$$R_{Pdp} = V_{Pdp}OP_{Pdp} + R_{PDbas} \qquad\qquad 15)$$

Figure 2 illustrates the sigmoidicity of the relationships that the above equations imply for the steady-state *per* expression rate *vs.* the levels of CLK and PER. The expression rate varies from a minimum of $R_{Pbas}$ (in the absence of CLK, or with high $[PER_{nuc}]$) to a maximum of $V_{Per} + R_{Pbas}$ (for saturating [CLK] with PER absent). Graphs of the steady-state *vri* and *Pdp-1* expression rates are very similar (not shown).

The known circadian regulators of *Clk* are VRI and PDP-1. Transcription of *Clk* was assigned a maximal rate $V_{clk}$. Repression by VRI was represented by multiplying $V_{clk}$ by a Hill function of [VRI]. At least five binding sites for VRI exist in the *Clk* promoter region and gene (Glossop et al., 2003). Multiple VRI binding sites suggest a Hill coefficient greater than 1 may be appropriate. A coefficient of 2 was used for the simulations presented in Figs. 3-7. A small basal transcription rate $R_{Cbas}$ was assigned to *Clk* in the limit of high [VRI]. Activation of *Clk* by PDP-1 was represented with a Hill function of [PDP-1], with a Hill coefficient of 2. The equation for the rate $R_{Clk}$ of *Clk* expression uses the product of the Hill functions for PDP-1 and VRI,

$$R_{Clk} = V_{Clk} \left( \frac{[PDP\text{-}1]^2}{[PDP\text{-}1]^2 + K_{PDC}^2} \right) \left( \frac{K_{VC}^2}{[VRI]^2 + K_{VC}^2} \right) + R_{Cbas} \qquad 16)$$

The PDP-1 and VRI DNA binding domains are similar. Competition between PDP-1 and VRI has been demonstrated at one PDP-1 binding site and is thought to take place at the other sites (Cyran et al., 2003). Eq. 16 does not represent the details of this competition. Instead, Eq. 16 was constructed to qualitatively represent nonlinear activation by PDP-1 and repression by VRI. We note that activators of *Clk* other than PDP-1 may also play a significant role, because $Clk^{Jrk}$ mutant *Drosophila* have low levels of PDP-1, but high levels of *Clk* mRNA (near the wild-type peak) (Cyran et al., 2003; Glossop et al., 1999).

To reduce the number of equations, *vri, per, Clk,* and *Pdp-1* mRNAs were not included as variables. Gene expression rates (Eqs. 13-16) were used to describe VRI, PDP-1, and CLK synthesis. First-order degradation was assumed. Division of VRI, PDP-1, and CLK between cytoplasm and nucleus was not considered. An additional, small first-order degradation rate constant $k_d$ of ~ 0.01 $hr^{-1}$ was assumed for every molecular species in the model. This practice is common (*e.g.*, Leloup et al., 1998) and ensures concentrations always remain bounded during simulations. These assumptions yield the following differential equations for the concentrations of CLK and VRI,

$$\frac{d[CLK]}{dt} = R_{Clk} - v_{dclk}[CLK] - k_d[CLK] \qquad 17)$$

$$\frac{d[VRI]}{dt} = R_{Vri} - v_{dvri}[VRI] - k_d[VRI] \qquad 18)$$

The differential equation for [PDP-1] is similar, but includes a delay of several hours between regulation of *Pdp-1* as given by Eq. 15 and regulation of PDP-1 synthesis. *Pdp-1* and *vri* are both activated by CLK, but during a circadian oscillation, the rise in [PDP-1] lags the rise in [VRI] by 3-4 hrs (Cyran et al., 2003). The mechanism underlying this delay is not known. To implement this delay, the rate of PDP-1 synthesis, $R_{Pdp}$ in Eq. 15, was continuously computed and stored. The stored values were used to calculate the rate of PDP-1 synthesis ~ 3 hrs later ($\tau_{delay}$ in Table I). The following equation describes these dynamics, with brackets denoting the discrete time delay,

$$\frac{d[\text{PDP-1}]}{dt} = \left\langle R_{\text{Pdp}} \right\rangle \tau_{delay} - v_{\text{dpdp}}[\text{PDP-1}]$$

$$- k_d[\text{PDP-1}]$$

19)

PER is progressively phosphorylated over a period of hours following its synthesis (Edery et al., 1994). The model assumes sequential phosphorylations, as have previous models (Smolen et al., 2001; Leloup and Goldbeter, 1998; Goldbeter, 1995). The number of phosphorylations and their distribution between cytosol and nucleus is not known. The model represents 4 phosphorylations of PER, 2 of which are cytosolic. Non-phosphorylated cytosolic PER is denoted $P0_{cyt}$, and its synthesis rate is given by Eq. 13. One phosphorylation gives $P1_{cyt}$, and a second yields $P2_{cyt}$. For simplicity, phosphorylations are assumed to be irreversible and described by Michaelis-Menten rate expressions. The Michaelis constant is denoted $K_{pcyt}$, and the maximal velocity is denoted $v_{pcyt}$. $P2_{cyt}$ is removed by transport into the nucleus, which is described by a Michaelis constant of $K_{pcyt}$ and a maximal velocity of $v_{trans}$. These assumptions yield the following differential equations for $[P0_{cyt}]$, $[P1_{cyt}]$, and $[P2_{cyt}]$,

$$\frac{d[P0_{cyt}]}{dt} = R_{Per} - v_{pcyt}\frac{[P0_{cyt}]}{K_{pcyt} + [P0_{cyt}]} - k_d[P0_{cyt}]$$

20)

$$\frac{d[P1_{cyt}]}{dt} = v_{pcyt}\frac{[P0_{cyt}]}{K_{pcyt} + [P0_{cyt}]}$$

$$- v_{pcyt}\frac{[P1_{cyt}]}{K_{pcyt} + [P1_{cyt}]} - k_d[P1_{cyt}]$$

21)

$$\frac{d[P2_{cyt}]}{dt} = v_{pcyt}\frac{[P1_{cyt}]}{K_{pcyt} + [P1_{cyt}]}$$

$$- v_{trans}\frac{[P2_{cyt}]}{K_{trans} + [P2_{cyt}]} - k_d[P2_{cyt}]$$

22)

Nuclear PER phosphorylation states are denoted $P0_{nuc}$, $P1_{nuc}$, and $P2_{nuc}$. $P0_{nuc}$ is formed from $P2_{cyt}$ by transport into the nucleus. The resulting differential equations for $[P0_{nuc}]$ and $[P1_{nuc}]$ are

$$\frac{d[P0_{nuc}]}{dt} = v_{trans}\frac{[P2_{cyt}]}{K_{trans} + [P2_{cyt}]}$$

$$- v_{pnuc}\frac{[P0_{nuc}]}{K_{pnuc} + [P0_{nuc}]} - k_d[P0_{nuc}]$$

23)

$$\frac{d[P1_{nuc}]}{dt} = v_{pnuc} \frac{[P0_{nuc}]}{K_{pnuc} + [P0_{nuc}]}$$

$$- v_{pnuc} \frac{[P1_{nuc}]}{K_{pnuc} + [P1_{nuc}]} - k_d[P1_{nuc}] \qquad 24)$$

$P2_{nuc}$ is formed by phosphorylation of $P1_{nuc}$. Experimentally, PER is seen to degrade relatively rapidly following its phosphorylation (Edery et al., 1994). In the model, degradation of $P2_{nuc}$ is described by a Michaelis-Menten term with maximal velocity $v_{degp}$ and Michaelis constant $K_{degp}$. The differential equation for $[P2_{nuc}]$ is therefore

$$\frac{d[P2_{nuc}]}{dt} = v_{pnuc} \frac{[P1_{nuc}]}{K_{pnuc} + [P1_{nuc}]}$$

$$- v_{degp} \frac{[P2_{nuc}]}{K_{degp} + [P2_{nuc}]} - k_d[P2_{nuc}] \qquad 25)$$

The relative peak-to-trough amplitude (per cent difference) of simulated circadian oscillations was found to be significantly enhanced if phosphorylation and degradation of nuclear PER were modeled as approximately zero-order processes. Therefore, the corresponding Michaelis constants ($K_{pnuc}$ and $K_{degp}$ in Eqs. 23-25) are assigned very low values (0.001 nM and 0.01 nM respectively).

Certain combinations of molecular species play a significant role in the dynamics of the model, or display time courses that can be compared with experimental data. The total concentration of nuclear PER is given as

$$[PER_{nuc}] = [P0_{nuc}] + [P1_{nuc}] + [P2_{nuc}] \qquad 26)$$

The model dynamics are fully determined by Eqs. 1-26.

The concentration of cytosolic PER is given by an expression analogous to Eq. 26

$$[PER_{cyt}] = [P0_{cyt}] + [P1_{cyt}] + [P2_{cyt}] \qquad 27)$$

Because concentrations are referenced to the total cell volume, the total concentration of PER, $[PER_{tot}]$, is the sum of the nuclear and cytosolic concentrations

$$[PER_{tot}] = [PER_{cyt}] + [PER_{nuc}] \qquad 28)$$

The above model does not represent expression of *tim* or the dynamics of the TIM gene product. Recent data suggests monomeric nuclear PER is the species responsible for inhibiting CLK's activation of *per*, *tim*, and other core circadian genes (Weber and Kay, 2003; Rothenfluh et al., 2000). Thus, TIM does not appear to directly regulate transcription. However, TIM indirectly regulates transcription by regulating the level and localization of PER. *Tim* null homozygotes are arrhythmic, and [PER] is decreased (Price et al., 1995). PER nuclear localization is blocked in homozygotes (Vosshall et al., 1994). PER and TIM heterodimerize, and it has been suggested that heterodimer formation suffices for PER nuclear entry (Saez and Young, 1996). However, Shafer et al. (2002) report PER accumulates in the nucleus of ventrolateral neurons during early night, while most TIM remains cytosolic. These data suggest a mechanism in which TIM interacts with an unidentified cytosolic factor and block its ability to retain PER in the cytosol, as first noted by Vosshall et al. (1994). Alternatively, PER-TIM heterodimerization might be required for nuclear entry, but during the early night, TIM may be re-exported (Shafer et al., 2002). Because the mechanism of TIM's facilitation of PER nuclear entry is uncertain, and because TIM appears not to directly regulate transcription, it appears reasonable not to explicitly model TIM dynamics. The number of equations is thereby reduced considerably. However, in future work we plan to consider the effect of cyclical modulation of PER nuclear entry by TIM.

For most simulations, a standard set of model parameter values was used. Table I divides the parameters into groups that describe similar processes and displays the standard values. As noted in the footnote to Table I, these values include a scaling factor $\lambda$ to adjust the free-running oscillation period to 24.0 hrs.

Data to constrain values of kinetic parameters are generally lacking. To obtain standard parameter values, it was necessary to rely on trial-and-error variation. Values were found that generated simulations similar to experimental data for the following phenomena: stable circadian oscillations robust to small parameter changes, simulation of entrainment to light pulses and simulation of a photic phase-response curve, and oscillations and steady states of circadian gene mutants. For example, in Table I, the maximal rates of gene expression ($V_{Clk}$, $V_{Vri}$, $V_{Per}$, and $V_{Pdp}$ in Eqs. 13-16) cover a broad range. These values were chosen *via* extensive efforts to simulate both normal circadian oscillations and steady states of mutants (Figs. 3A, 4A-C).

The model does not incorporate post-transcriptional regulation of the stability or translation of *per* or *Clk* mRNAs. Such regulation has not yet been well characterized, although it appears to exist (Stanewsky et al., 2002; Stanewsky et al., 1997; So and Rosbash, 1997). The model also does not incorporate the post-transcriptional regulation of [CLK] (Kim et al., 2002) because the function of this regulation is not yet known.

**Numerical methods.**

The forward-Euler method was used for integration of differential equations. Integration time steps were reduced until no significant difference was seen upon further reduction ($\sim 2 \times 10^{-4}$ hr). The model was programmed in Java and simulated on a Pentium 3 microcomputer. Simulation programs are available from the authors upon request.

## Results

**Simulated oscillations of *Drosophila* core gene product levels resemble experimental oscillations.**

Figure 3A illustrates that the model of Fig.1 simulates large-amplitude circadian oscillations in the concentrations of CLK, VRI, PDP-1, and total PER (cytosolic plus nuclear, [$PER_{tot}$] in Eq. 28) with a period of 24 hrs. Effects of light are not simulated, so these oscillations correspond to a free-running rhythm in constant darkness (DD). For [$PER_{tot}$], the peak and average are ~2-fold higher than for [CLK]. The peak of [VRI] is also higher than the peak of [$PER_{tot}$], and the [VRI] peak leads the [$PER_{tot}$] peak by ~5 hrs. In these respects, the simulation resembles experimental data (Glossop et al., 2003; Bae et al., 2000; Lee et al., 1998). The delay between the accumulation of [VRI] and that of [$PER_{tot}$] is due to the time required for the multiple phosphorylation states of PER to accumulate. Figure 3A displays a lag of ~3 hrs between the rise of [VRI] and that of [PDP-1]. A similar lag is seen experimentally (Cyran et al., 2003). Because CLK activates both *vri* and *Pdp-1* expression, this lag could only be simulated by introducing a time delay ($\tau_{delay}$ in Eq. 19) between CLK's regulation of *Pdp-1* expression and PDP-1 protein accumulation.

Figure 3B illustrates oscillations in nuclear [PER] concentration ([$PER_{nuc}$]), which is assumed to mediate repression of CLK-driven transcription (Eqs. 1-3). Figure 3B also illustrates oscillations of newly synthesized cytosolic PER ([$P0_{cyt}$]), and doubly phosphorylated cytosolic PER ([$P2_{cyt}$]). There is a substantial delay (~ 6 hrs) between the peak of [$P0_{cyt}$] and that of [$PER_{nuc}$], due to the time required for cytosolic PER to undergo two phosphorylation events and then to be transported into the nucleus (Eqs. 20-23). After PER enters the nucleus, it is assumed to undergo two more phosphorylations and then to be degraded (Eqs. 23-25). Similar dynamics are seen experimentally; multiple phosphorylations of PER over hours are followed by rapid degradation of phosphorylated PER (Edery et al., 1994).

Although *per* mRNA is not an explicit model variable, comparison of the time courses of [$PER_{tot}$] (Fig. 3A) and of [$P0_{cyt}$] (Fig. 3B) illustrates that the model represents much of the 5-6 hr delay observed between the time courses of *per* mRNA and [$PER_{tot}$] (Glossop et al., 1999; Vosshall et al., 1994; Hardin et al., 1990). The time required for PER phosphorylation and nuclear entry results in a delay of ~ 5 hrs between the peak of the simulated time course of the initial form of PER, [$P0_{cyt}$] (at $t = 11$ hrs), and the peak of [$PER_{tot}$] (at $t = 16$ hrs). A similar delay is seen between the initial rise of [$P0_{cyt}$] and the rise of [$PER_{tot}$].

The mechanism underlying the oscillations in Fig. 3A, B can be summarized as follows. At time $t = 4$ hrs, [CLK] is rising close to its peak level, and [$PER_{tot}$] and [$PER_{nuc}$] are falling. These changes activate histone acetylation at the *vri, per,* and *Pdp-1* promoters (Eqs 1-3) and transcription of *vri, per,* and *Pdp-1* (Eqs. 13-15). [VRI] begins to rise at $t \sim 7$ hrs. The continuing decay of nuclear PER from the previous cycle of PER accumulation prevents [$PER_{tot}$] from rising until $t \sim 10$ hrs, and the delay $\tau_{delay}$ between *Pdp-1* transcription and PDP-1 synthesis prevents [PDP-1] from rising until $t \sim 11$ hrs.

[PER$_{nuc}$] rises slightly after [PER$_{tot}$] (Fig. 3B). From $t \sim 9$ hrs to $t \sim 17$ hrs, [VRI] is high, *Clk* is repressed, and [CLK] is falling. Elevated [PER$_{nuc}$] promotes histone deacetylation by decreasing the acetylation rate constants (Eqs. 1-3). Therefore *vri, per,* and *Pdp-1* transcription are inhibited after $t \sim 13$ hrs, when [PER$_{nuc}$] is high. Because of this inhibition, [VRI], [PER$_{tot}$], [PER$_{nuc}$], and [PDP-1] pass in sequence through their peaks during $t \sim 12 - 16$ hrs and then decline. By $t \sim 24$ hrs [VRI], [PER$_{tot}$], and [PDP-1] have dropped well below their peak levels. The fall in [VRI] has removed its inhibition of *Clk*, and [CLK] is therefore rising again, to begin another cycle.

There is considerable variation in standard values for the basal synthesis rates of CLK, PER, VRI, and PDP-1 (respectively R$_{Cbas}$, R$_{Pbas}$, R$_{Vbas}$, and R$_{PDbas}$ in Table I). The values vary between R$_{Cbas}$ = 0.001 nM hr$^{-1}$ and R$_{PDbas}$ = 0.36 nM hr$^{-1}$. We found empirically that, for simulating circadian oscillations (Fig. 3A) a small R$_{Cbas}$ was required to make the trough of [CLK] oscillations near zero. This, in turn, was required to sustain large peak-to-trough ratios of oscillations in [PER$_{tot}$], [VRI], and [PDP-1]. In contrast, the peak of [CLK] occurs just prior to the trough of [PDP-1] (Fig. 3A). To maintain substantial CLK synthesis and a significant [CLK] peak, it was necessary that the [PDP-1] trough not be too low. Therefore R$_{PDbas}$ had to be relatively high.

As discussed in Model Development and Numerical Methods, activation of *vri, per,* and *Pdp-1* expression is proportional to the variables OP$_{Vri}$, OP$_{Per}$, and OP$_{Pdp}$, which represent promoter accessibilities to RNA polymerase (Eqs. 7-15). A putative requirement for acetylation of multiple histones is represented by a sigmoidal, "threshold" dependence of these promoter accessibilities on the variables AC$_{Vri}$, AC$_{Per}$, and AC$_{Pdp}$, which represent acetylation of single histones. The promoter accessibilities relax to steady-state values given by 5$^{th}$ powers of AC$_{Vri}$, AC$_{Per}$, and AC$_{Pdp}$ (Eqs. 7-9). Simulations suggested this sigmoidicity is essential. If only 1$^{st}$ powers of AC$_{Vri}$, AC$_{Per}$, and AC$_{Pdp}$ were used in Eqs. 7-9, circadian oscillations could not be simulated irrespective of the values for the parameters in Table I. Oscillations could be sustained if 3$^{rd}$ or 4$^{th}$ powers were used. However, Fig. 3C illustrates that with 3$^{rd}$ powers, the peak-to-trough ratio of oscillations is considerably reduced. Experimental time courses (Cyran et al., 2003; Glossop et al., 2003; Bae et al., 2000; Lee et al., 1998) exhibit much larger peak-to-trough ratios, more similar to Fig. 3A.

Circadian rhythms of histone acetylation at mammalian *per1* and *per2* promoters appear synchronous with the rhythms of *per1* and *per2* mRNA (Etchegaray et al., 2003, their Fig. 1). Do our simulated oscillations display similar synchrony? Our model does not include mRNA concentrations. However, time courses of the *per, vri,* and *Pdp-1* transcription rates (R$_{Per}$, R$_{Vri}$, and R$_{Pdp}$, Eqs. 13-15) can be compared with time courses of acetylation at the *per, vri,* and *Pdp-1* promoters (AC$_{Vri}$, AC$_{Per}$, and AC$_{Pdp}$, Eqs. 4-6). For the oscillations of Fig. 3, the comparison reveals very little lag ($\sim$1½ hours) between the peaks of acetylation and the following peaks of transcription rates (not shown). Therefore, this simulation appears qualitatively consistent with the data of Etchegaray et al. (2003).

The model describes *Clk* activation by PDP-1, and *Clk* repression by nuclear PER, with nonlinear functions. In Eq. 16, a Hill coefficient of 2 is assumed to describe both these processes. We repeated the simulation of Fig. 3A with these nonlinearites diminished (Hill coefficients of 1 in Eq. 16) or enhanced (Hill coefficients of 3 in Eq. 16). The effects of these changes were minor. Oscillations were preserved, with circadian periods (23.8 and 24.3 hrs respectively). The relative phases and amplitudes of [PER], [CLK], [PDP-1], and [VRI] oscillations remained similar to those in Fig. 3A. Peak-to-trough ratios were slightly reduced with Hill coefficients of 1 (not shown).

**Simulated oscillations are robust to parameter variations.**

Despite heterogeneity between individuals in the values of parameters describing gene expression regulation, individual *Drosophila* maintain very similar circadian periods in DD. For example, Bao et al. (2001) reported periods of $24.3 \pm 0.06$ hrs for 43 wild-type flies. A model of circadian rhythm generation should therefore be robust in the sense that small parameter variations should not result in large period alterations. Robustness was tested with a method used in previous investigations (Smolen et al., 2001; Lema et al., 2000; Goldbeter, 1995). Each parameter was increased or decreased by a moderate amount (20%) from its standard value in Table I. The effects on oscillation amplitude and period were determined, with all other parameters kept at standard values. There are 38 model parameters (Table I, not including $k_{light}$ or N). Therefore, 77 simulations were carried out including the control with standard parameter values. Figure 3D plots the period and amplitude of the resulting oscillations. The amplitude was measured as the peak-to-trough difference in [$PER_{tot}$].

Oscillations were preserved in all simulations. The mechanism of oscillation remained as described for control oscillations (Figs. 3A-B). Most oscillations have periods and amplitudes close to control values (24.0 hrs, 3.30 nM). Three parameter changes yielded periods differing more than 3 hours from control. The largest period and amplitude decreases (20.3 hrs, 1.54 nM) occurred for the 20% decrease in $v_{pnuc}$. A 20% increase in $v_{pnuc}$ produced the largest amplitude increase (5.50 nM) and a significant period increase (27.5 hrs). Increasing $v_{pnuc}$ acts to inhibit and delay accumulation of $PER_{nuc}$ by promoting the formation of the most highly phosphorylated species of PER, $P2_{nuc}$, which is rapidly degraded. The inhibition of $PER_{nuc}$ accumulation enhances CLK's activation of *per, vri,* and *Pdp-1* expression. Therefore, [$PER_{tot}$] reaches a higher peak and takes longer to degrade. The largest period increase (28.3 hrs) occurred for a 20% decrease in $K_{PP}$ (Eq. 2). Perhaps uncharacterized mechanisms exist in *Drosophila* to reduce the sensitivity of oscillations to $v_{pnuc}$ and $K_{PP}$. Overall, the lack of large period or amplitude changes suggests the model is sufficiently robust to be regarded as a reasonable representation of biochemical mechanisms responsible for circadian oscillations in *Drosophila*.

**Simulations of mutations in core clock genes, and of enhanced PER phosphorylation, are qualitatively similar to experimental data**.

Starting from standard parameter values (Table I, Fig. 3A-B), the model was perturbed to simulate the effects of homozygous null mutations in *per* and *Clk*. Following established nomenclature, these mutations are denoted *per$^{01}$* and *Clk$^{Jrk}$*. In these mutations, nonfunctional proteins appear to be produced. Therefore, to simulate *Clk$^{Jrk}$*, the transcriptional activation terms for *per, Pdp-1,* and *vri* in Eqs. 1-3 (the Hill functions of [CLK]) were set to zero. To simulate *per$^{01}$*, the transcriptional repression terms in Eqs.

1-3 (the Hill functions of [PER$_{nuc}$]) were set to one. The $per^{01}$ ; $Clk^{Jrk}$ double mutant was also simulated, by applying both the $Clk^{Jrk}$ and $per^{01}$ perturbations.

Figures 4A-B illustrate simulations of $per^{01}$ and $Clk^{Jrk}$. Levels of [PER$_{tot}$], [VRI], [PDP-1], and [CLK] can be compared with simulated wild-type (WT) oscillations (Fig. 3A). For $per^{01}$, the PER$_{tot}$ level is about one-third of the WT peak, qualitatively consistent with observations that $per$ mRNA levels are ~ 50 % of the WT peak (So and Rosbash, 1997; Hardin et al., 1990). [CLK] is very low, qualitatively consistent with the observation of a low $Clk$ mRNA level in $per^{01}$ (about 20% of the WT peak, Glossop et al., 1999). However, [VRI] is nevertheless high, near its peak in Fig. 3A. This apparent contradiction results because the low [PER$_{tot}$] and low [PER$_{nuc}$] result in little repression of $vri$ histone acetylation and $vri$ expression (Eqs. 1, 16). Under this condition, even low [CLK] suffices to drive substantial VRI synthesis. This simulation does show some discrepancy with data, in that experimentally, $vri$ mRNA is at intermediate, not peak, levels in $per^{01}$ flies (Blau and Young, 1999).

In the $Clk^{Jrk}$ simulation (Fig. 4B), PER$_{tot}$ is very low. This is qualitatively consistent with the observation that in $Clk^{Jrk}$ flies, $per$ mRNA levels are very low (< 15% of the WT peak, Allada et al., 1998). In contrast, the level of [CLK] is similar to the peak of [CLK] in Fig. 3A. This is consistent with the observation that in $Clk^{Jrk}$ flies, $Clk$ mRNA is steady and near the WT peak (Glossop et al., 1999). Because the non-functional CLK does not activate $vri$, [VRI] is low (near the trough of [VRI] in Fig. 3A). Because VRI is not repressing $Clk$, a relatively high [CLK] is maintained even though the level of $Clk$'s activator, PDP-1, is relatively low. In $per^{01}$ ; $Clk^{Jrk}$ double mutants, $Clk$ mRNA is also observed to be near the WT peak (Glossop et al., 1999). The $per^{01}$ ; $Clk^{Jrk}$ simulation yielded a high [CLK] level, identical to $Clk^{Jrk}$. This result was expected, because after the Hill functions of [CLK] in Eqs. 1-3 are set to zero to simulate $Clk^{Jrk}$, no further change occurs if the Hill functions of [PER$_{nuc}$] are set to one to simulate $per^{01}$.

If any of the multiple phosphorylations of PER are obligatory for nuclear entry of PER, then these phosphorylations constitute kinetic steps within the negative feedback loop in which PER acts to repress $per$ expression. In the model, both cytosolic phosphorylations of PER lie within this feedback loop. Therefore, increasing the PER phosphorylation rate is predicted to speed up the dynamics of this loop and shorten the oscillation period. Fig. 4C confirms this prediction. A 65% increase was applied to the maximal phosphorylation velocities of PER and to the velocity of PER nuclear transfer ($v_{pcyt}$, $v_{pnuc}$, and $v_{trans}$, Eqs. 20-25). The period decreased to 18.6 hrs. The increase in $v_{trans}$ was necessary to reduce the period below 19 hrs. It is plausible that PER nuclear transfer involves a phosphorylation, so that its rate would be affected by altering PER phosphorylation kinetics. In Fig. 4C, [PDP-1] and [VRI] oscillations are larger than control oscillations (Fig. 3A). This occurs because the increased $v_{pnuc}$ allows more rapid conversion of nuclear PER into fully phosphorylated P2$_{nuc}$. P2$_{nuc}$ rapidly degrades, so PER$_{nuc}$ accumulation is inhibited. Therefore, the expression of $vri$ and $Pdp-1$ is not as repressed by PER$_{nuc}$, yielding higher peaks of [PDP-1] and [VRI].

The $per^{S}$ mutation shortens the free-running circadian period to ~ 19 hrs (Konopka and Benzer, 1971) and results in more rapid PER phosphorylation, with a greater proportion of PER$^{S}$ extensively phosphorylated at earlier Zeitgeber time (Edery et al., 1994). The simulation of Fig. 4C may qualitatively represent these aspects of the $per^S$ phenotype. Another mutation, $dbt^S$, also shortens the free-running circadian period (Price et al., 1998). The $dbt$ gene product (DBT) is a casein kinase Iε homologue that phosphorylates PER (Kloss et al., 1998). Its activity may be increased in $dbt^S$, but this has not been biochemically confirmed. The period in Fig. 4C is similar to that of $dbt^S$ (~18 hrs, Price et al., 1998). Two other mutations, $dbt^g$ and $dbt^h$, increase the free-running period and may decrease DBT activity (Suri et al., 2000). However, if these mutations are simulated by a parameter change opposite to that in Fig. 4C (decreasing $v_{pcyt}$, $v_{pnuc}$, and $v_{trans}$), the period decreases. Therefore, the model may not satisfactorily simulate $dbt^g$ and $dbt^h$ mutations, although a 32% decrease in $v_{pcyt}$ alone does yield a period of 29 hrs, similar to a $dbt^h$ homozygote or $dbt^g$ heterozygote.

Blau and Young (1999) reported that circadian rhythms are preserved in flies having one functional copy of $vri$ (i.e., $vri^{null}/+$), but that the period was decreased by 0.4 − 0.8 hrs. In Fig. 5A, a $vri^{null}/+$ mutation was modeled with a 50% decrease in the maximal and basal velocities of $vri$ transcription ($V_{Vri}$ and $R_{Vbas}$, Eq. 14). The [VRI] and [PER$_{tot}$] oscillations are similar to WT oscillations (Fig. 3A) except that the amplitude of the [VRI] oscillation is diminished. Repression of CLK synthesis is thereby relieved and [CLK] levels are above the WT control (Fig. 3A). The period is decreased to 23.3 hrs, 0.7 hrs below WT. We also simulated heterozygous null mutations of $Clk$, $Pdp-1$, and $per$ ($Clk^{Jrk}/+$, $Pdp^{null}/+$, $per^{01}/+$) with 50% decreases in the corresponding maximal and basal transcription rates in Eqs. 16, 15, and 13 ($V_{Clk}$, $R_{Cbas}$, $V_{Pdp}$, $R_{PDbas}$, $V_{Per}$, $R_{Pbas}$). Experimental periods of these heterozygotes in DD are similar to WT: 24.8 hrs for $Clk^{Jrk}/+$ (Allada et al., 1998), 23.6 hrs for $Pdp^{null}/+$ (Cyran et al., 2003), and 25.2 hrs for $per^{01}/+$ (Konopka and Benzer, 1971). Simulated $Clk^{Jrk}/+$ and $Pdp^{null}/+$ periods were close to the experimental periods: 25.0 hrs for $Clk^{Jrk}/+$, 23.9 hrs for $Pdp^{null}/+$. The appearance of the oscillations was similar to WT (Fig. 3A), except the [CLK] and [PDP-1] amplitudes were respectively decreased by ~ 50% in $Clk^{Jrk}/+$ and $Pdp^{null}/+$. In contrast, the $per^{01}/+$ simulation failed to sustain oscillations, contradicting experimental data. A steady state was obtained, with [VRI] high (7.0 nM) and [CLK] and [PER$_{tot}$] very low.

Because decreased $vri$ transcription lowered the period and increased the average of [CLK] (Fig. 5A), we predicted that simulating $Clk$ overexpression would also decrease the period. Figure 5B verifies this prediction. For this simulation, activation of $Clk$ by PDP-1 and repression by PER was preserved, i.e., Eq. 16 was used. However, a constant value of 0.3 nM hr$^{-1}$ was added to the CLK synthesis rate $R_{clk}$ (Eqs. 16-17). The shapes and relative phases of the oscillations in [PER$_{tot}$], [PDP-1], and [VRI] are not significantly affected by this increase in CLK synthesis. The period is decreased to 23.4 hrs, 0.6 hrs below control (Fig. 3A).

Hao et al. (1999) rescued circadian rhythmicity in $per^{01}$ mutants by introducing a $per$ transgene with a promoter more active than that of WT $per$. However, the rhythm had a short period (22.5 hrs). The model simulates a similar period decrease with $per$ ovexpression. If the maximal and basal rates of $per$ expression ($v_{per}$ and $R_{Pbas}$) are increased by 60%, the period decreases to 22.4 hrs.

**Removal of positive feedback eliminates circadian expression of some clock genes, but not of *per*. Negative feedback is essential for oscillations of all clock genes.**

The positive feedback loop relying on repression of *Clk* was elucidated by Glossop et al. (1999) and is summarized in the legend of Fig. 1. Constitutive *vri* or *Clk* expression can be simulated by fixing [VRI] or [CLK]. Either manipulation eliminates the positive feedback loop by eliminating dynamic repression of *Clk*. In previous models, fixation of [CLK] did not prevent circadian oscillations of *per* expression (Smolen et al., 2002; Smolen et al., 2001). Similarly, Fig. 6A illustrates that when [CLK] is fixed at half of its peak in Fig. 3A oscillations in [PER$_{tot}$] persist, with period near circadian (26.0 hrs). The oscillations in [PER$_{tot}$] and [PER$_{nuc}$] result in peaks of PER$_{nuc}$ that cyclically inhibit CLK's ability to activate *Pdp-1, vri* and *per*. Therefore, [PDP-1] and [VRI] also oscillate.

When positive feedback was removed by fixing [VRI], Fig. 6B illustrates that [PER$_{tot}$] and [PDP-1] continued to oscillate with a period qualitatively near to circadian (19.7 hrs). Oscillations in [PDP-1] drive oscillations in PDP-1's activation of *Clk* expression (Eq. 16). Low-amplitude oscillations in [CLK] result. Fixing [VRI] also removes the negative feedback loop in which CLK activates *vri* and VRI represses *Clk*. Therefore, Fig. 6B suggests that this negative feedback loop is not essential for circadian oscillations of *per* or *pdp-1* expression, although it modulates the amplitude of [CLK] and [VRI] oscillations.

Experimentally, *vri* overexpression often leads to arrhythmicity (Blau and Young, 1999). In the model, maintaining [VRI] at a high level strongly inhibits circadian oscillations. Fixing [VRI] 60% above its level in Fig. 6B yields an amplitude of only 0.85 nM for [PER$_{tot}$]. The period is decreased to 14.4 hrs. This decrease is discrepant with data, in that Blau and Young (1999) found that the free-running period of rhythmic *Drosophila* with *vri* overexpressed was lengthened by ~ 1.5 – 3 hrs. In the model, higher [VRI] abolishes oscillations. In contrast, the dynamics in Fig. 6A are relatively insensitive to variations in [CLK]. [PDP-1] and [PER$_{tot}$] oscillations of nearly circadian period are preserved when [CLK] is fixed at any value between 0.1 nM and infinity, with "infinite [CLK]" implemented in Eqs. 1-3 by setting the functions of [CLK] on the right-hand sides to 1.0. Eqs. 1-3 therefore qualitatively model the following situation: Even with very high [CLK], only a limited and saturated amount of [CLK] is bound at E-box sites to promote histone acetylation. PER$_{nuc}$ still binds cyclically to CLK – E-box complexes, repressing CLK's action and sustaining oscillations of acetylation and gene expression. Circadian oscillations of [PDP-1] and [PER$_{tot}$] are also simulated if [VRI] is fixed at a low value or at zero (not shown).

The model contains a second positive feedback loop in which PDP-1 activates *Clk* and CLK reciprocally activates *Pdp-1* (Fig. 1A). To remove only this feedback loop, *Pdp-1* expression was fixed by holding [PDP-1] constant. Oscillations of [VRI], [CLK], and [PER$_{tot}$] were preserved. The shapes and relative phases were similar to the control oscillations (Fig. 3A) for a wide range of fixed [PDP-1] (from < 0.3 nM to arbitrarily high levels). The period remained circadian.

The above simulations suggest neither of the positive feedback loops in Fig. 1A may be required to sustain circadian oscillations in the levels of PER, VRI, or free CLK not complexed with PER. What about the core negative feedback loop in which PER represses *per* transcription by binding CLK? Simulations were carried out with this loop removed by fixing the PER synthesis rate ($R_{Per}$, Eqs. 13 and 20). Under this condition, oscillations were abolished (not shown). Only steady states of all concentrations were obtained, irrespective of the values chosen for the parameters in Table I.

It can be hypothesized that, although the positive feedback loops in Fig. 1A are not required for circadian oscillations *per se*, they may increase the robustness of oscillation amplitude and period to modest variations in parameters. To determine whether our model supports this hypothesis, we constructed scatter plots of oscillation period *vs.* [PER_tot] oscillation amplitude, analogous to Fig. 3C. Plots were constructed for three conditions: 1) [VRI] fixed at 3.0 nM as in Fig. 6B, thereby removing the PER−CLK positive feedback loop and the VRI−CLK negative feedback loop; 2) [CLK] fixed at 0.79 nM as in Fig. 6A, thereby eliminating all feedback loops except the PER−CLK negative feedback loop; and 3) [PDP-1] fixed at 3.0 nM, thereby eliminating only the CLK−PDP-1 positive feedback loop. To construct each scatter plot, each model parameter was increased or decreased by 20% from its standard value. There are 38 model parameters. Therefore, 76 simulations were carried out for each scatter plot. We found that none of the plots showed significantly less robustness to parameter variation than did the control plot with all feedback loops present (Fig. 3C). The per cent difference between minimal and maximal amplitudes or periods in each plot was not greater than in Fig. 3C, with the single exception that decreasing $v_{pnuc}$ by 20% abolished oscillations for fixed [CLK]. Therefore, these simulations fail to support the hypothesis that either the positive feedback loops in Fig. 1A, or the VRI−CLK negative feedback loop, significantly increase the robustness of oscillations to moderate parameter variations.

**The model simulates photic entrainment and a phase-response curve similar to experimental curves.**

In *Drosophila*, light enhances degradation of phosphorylated TIM (Myers et al., 1996; Zeng et al., 1996). When TIM is removed from the complex of PER and TIM, phosphorylation of PER is strongly enhanced (Kloss et al., 2001). It is plausible that accelerated degradation of highly phosphorylated PER would result, and experiments have demonstrated light-induced enhancement of PER phosphorylation and degradation (Lee et al., 1996). Therefore, we modeled light exposure by adding a first-order degradation term to the right-hand side of each differential equation for a form of cytosolic PER. The degradation rate constant was denoted $k_{light}$. For example, Eq. 21 for [P1_cyt] has a term $-k_{light}[P1_{cyt}]$ added.

Figure 7A illustrates entrainment of the oscillations of Fig. 3A to a circadian light-dark (LD) cycle. The LD cycle was divided into equal light and dark portions and the period was set to 22 hrs. During the light phase, $k_{light}$ was set to 0.685 hr$^{-1}$ for P0_cyt, P1_cyt, and P2_cyt. The light phase occurs during the falling portion of the PER time course, as it does experimentally (Lee et al., 1998). Oscillations entrained to a 24-hr LD cycle look very similar to Fig. 7A (not shown). The amplitudes of the oscillations in PER and the other species are not significantly greater than in simulated DD conditions (Fig. 3A).

The entrainment range appears reasonably broad. For parameters as in Fig. 7A, entrainment occurs for an LD cycle length of 21 – 25 hrs.

A photic phase-response curve (PRC) was also constructed. Simulated light pulses were applied at 40 evenly spaced intervals during the circadian cycle in DD (Fig. 3A, B). During the pulse duration, $k_{light}$ was set to 0.685 $hr^{-1}$ for $P0_{cyt}$, $P1_{cyt}$, and $P2_{cyt}$. The pulse duration chosen was 2 hrs. In previous models, durations of 1-3 hrs have been used for degradation of TIM in response to light pulses (Leloup et al., 1998) or for degradation of an unspecified clock protein (Lema et al., 2000). Data illustrating PER disappearance subsequent to a light pulse lack sufficient temporal resolution to clearly determine the duration of increased PER degradation (Lee et al., 1996). These data also illustrate that under LD conditions, light pulses during the dark phase do not always accelerate PER disappearance. At Zeitgeber time 15, light pulses delay PER disappearance (Lee et al., 1996). Because the mechanism of this delay is not understood, and because PER regulation during LD cycles may differ significantly from that in DD, we have not modeled this additional complexity.

The phase advance or delay was determined seven cycles after each simulated light pulse (sufficient time for any transient dynamics to disappear). Figure 7B illustrates the resulting PRC (solid curve). Circadian time (CT) zero was chosen as coinciding with the peak of $PER_{tot}$ during the unperturbed oscillation (Shafer et al., 2002). By the classification of PRCs given in Winfree (1987), this PRC is Type 1 (average slope of 0 when plotted as in Fig. 7B). Figure 7B also illustrates an experimental PRC for *Drosophila* locomotor activity (data from Fig. 5 of Konopka et al., 1991). The agreement between the simulated and experimental PRCs appears quite good. The PRCs are similar in the magnitude of advances and delays, the number of hours of CT that correspond to advance *vs.* delay, and the steepness of the crossover from delay to advance. The PRCs both have a "dead zone" of approximately zero phase shift at CT 5-9. The crossover from delay to advance occurs at CT 17 (simulated) and CT 18 (experimental). The experimental PRC of Matsumoto et al. (1994) is similar, with a dead zone at CT 5-9 and a steep crossover from delay to advance at CT 19.

Increases in the strength of the simulated light pulse (*i.e.* in $k_{light}$) continued to yield a Type 1 PRC. However, Saunders et. al. (1994) illustrate Type 0 PRCs for *Drosophila melanogaster*. These PRCs were obtained with light pulses of very long duration (6 hrs). We therefore simulated very long-lasting increases in $k_{light}$ and examined whether a Type 0 PRC could result. A duration of 8 hrs was used, along with values of $k_{light}$ as high as 2.0 $hr^{-1}$. The PRC remained Type I. In contrast, if light was assumed to degrade nuclear as well as cytosolic PER, Type 0 PRCs could be obtained with strong stimuli (*e.g.*, $k_{light}$ = 2.0 $hr^{-1}$ for 3 hrs). With weaker stimuli, Type I PRCs were also obtained, but the shape of these PRCs failed to resemble experimental PRCs.

**Temperature compensation of oscillation period can be simulated.**

A well-known characteristic of circadian oscillators is temperature compensation. At different ambient temperatures, the constant-darkness period of circadian oscillations remains virtually constant in *Drosophila* (Pittendrigh et al., 1973; Pittendrigh, 1954). This finding is *a priori* unexpected, because a temperature increase tends to increase the rate of each biochemical reaction. The temperature dependence of a reaction is commonly described by the $Q_{10}$ factor, which equals the ratio of the rate 10° C above a reference temperature to the rate at the reference temperature. $Q_{10}$'s of ~ 2-3 are common for biochemical reactions (Segel, 1975). Since the period of a circadian oscillation is a function of the rates of many individual reactions, the period would be expected *a priori* to decrease as temperature increases.

One possible mechanism for temperature compensation is as follows. Processes such as nuclear transport of PER are composed of several elementary reactions (*e.g.*, binding of protein to nuclear pore complexes, conformation changes of nuclear pore complexes). The rate of elementary reactions that hinder PER transport, such as dissociation of PER from nuclear pore complexes, might increase more rapidly with temperature than do the rates of reactions that favor PER transport. In this case a temperature increase could decrease the rate of PER nuclear transport, and a $Q_{10}$ less than 1 would describe this decrease. Inhibiting PER nuclear transport would delay PER's repression of CLK-mediated transcription and lengthen the oscillation period (Hong and Tyson, 1997; Leloup and Goldbeter, 1997). This mechanism for temperature compensation is sometimes considered unsatisfying because it requires "fine-tuning" $Q_{10}$'s for separate kinetic processes. However, molecular evolution might have selected for phenotypes with a period resistant to moderate temperature changes, thereby resulting in fine-tuning of kinetic parameters.

Simulations suggest the model can qualitatively represent temperature compensation resulting from such fine-tuning. The control simulation of free-running oscillations (Fig. 3A) was used as the starting point. Initially, a $Q_{10}$ of 2.0 was applied to every kinetic process. This corresponds to multiplying the scaling factor λ in Table I by the $Q_{10}$, doubling every rate constant and maximal velocity and halving each time constant or delay. The oscillation period was halved to 12.0 hrs. Then, the simulation was repeated assuming that a temperature increase inhibited PER nuclear transport. A $Q_{10}$ of 0.62 was used to scale the maximal transport velocity $v_{trans}$. The period returned to circadian (24.2 hrs). Figure 7C displays these oscillations. Inhibition of PER nuclear transport tends to reduce nuclear PER levels, thereby reducing the repression of *Pdp-1* and *vri* by [$PER_{nuc}$]. As a result, [VRI] and [PDP-1] reach higher peaks in Fig. 7C than in the control simulation (Fig. 3A). Temperature compensation could also be simulated if a temperature increase was assumed to slow down cytosolic PER phosphorylation. In one simulation, a $Q_{10}$ of 2.0 was applied to all processes, except that a $Q_{10}$ of 0.83 scaled the cytosolic PER phosphorylation velocity $v_{pcyt}$. Temperature compensation was obtained, with a period of 24.2 hrs.

Temperature compensation *in vivo* probably involves additional mechanisms. For example, in the fungus *Neurospora crassa*, Rensing et al. (1997) found that temperature elevation initially depressed the expression rates of multiple genes, but these rates returned to normal in ~ 10 hours. An increase in the expression of heat shock proteins appeared to be involved in this homeostatic mechanism, which was suggested to contribute to temperature compensation. Because of the lack of data to delineate and constrain the *Drosophila* temperature compensation mechanism, we have not attempted to extend our model to simulate differences in temperature compensation between WT flies and $per^S$ mutants (Konopka et al., 1989) or $per^L$ mutants (Konopka et al., 1989; Curtin et al., 1995).

## Discussion

The present model was developed to represent the regulation of core clock component genes in *Drosophila* (*per, vri, Pdp-1,* and *Clk*). The model illustrates the ways in which negative and positive feedback loops (Fig. 1A) cooperate to generate oscillations of gene expression. The relative amplitudes and phases of simulated oscillations (Fig. 3A, B) resemble empirical data, except that the time course of CLK lags that of PER by ~12 hrs, whereas data exhibit a smaller lag of ~5 hrs (Lee et al., 1998). The smaller experimental lag could result from processes that delay the PER peak but are not represented in the model, such as additional phosphorylations of PER, or the observed delay of ~3 hrs between the rise of the *per* transcription rate and of *per* mRNA (So and Rosbash, 1997).

Simulated oscillations are robust to modest (20%) variations in parameters (Fig. 3D). Simulations of null mutations in *per* and *Clk* (Figs. 4A, B) yield steady states of gene product levels similar to experimental data, except for the high [VRI] in simulated $per^{01}$. The period of a simulated *vri* null heterozygote is decreased below control to 23.3 hrs (Fig. 5A), and the period also decreases when *per* overexpression is simulated. Similar decreases are seen experimentally (Blau and Young, 1999; Hao et al., 1999). The model also simulates observed effects of *Clk* and *Pdp-1* heterozygous null mutations on oscillation period. However, it fails to simulate the preservation of circadian rhythmicity with a $per^{01}$ heterozygous null mutation (Konopka and Benzer, 1971). Instead, the simulated decrease in PER transcriptional repression yields a steady state of sustained high VRI and PDP-1 levels. *In vivo*, an uncharacterized mechanism might compensate for the decrease in PER transcriptional repression in $per^{01}$ heterozygotes.

The negative feedback loop in which PER represses *per* by interacting with CLK is essential for simulation of circadian oscillations, as was the case in previous models (Smolen et al., 2001; Gonze et al., 2000; Leloup et al., 1998). Only this feedback loop is necessary for simulation of *per* expression oscillations. Two positive feedback loops exist. The first relies on repression of *Clk* and subsumes both negative feedback loops (legend to Fig. 1A) (Cyran et al., 2003; Glossop et al., 1999). Oscillations of *per, vri,* and *Pdp-1* expression are preserved when this positive feedback loop is removed by fixing *Clk* expression (Fig. 6A). Oscillation amplitudes of [PER], [VRI], and [PDP-1] are not reduced (compare Fig. 6A with Fig. 3A). Thus, these simulations do not suggest that positive feedback increases the amplitude of circadian oscillations.

The positive feedback loop relying on repression of *Clk* is necessary to drive oscillations of *Clk* expression, which regulate expression of clock-controlled genes (CCGs) outside the core oscillator. Microarray experiments have identified over 100 CCGs in *Drosophila* (Ceriani et al., 2002; Ueda et al., 2002; Claridge-Chang et al., 2001). The majority appear regulated by CLK (McDonald and Rosbash, 2001). Therefore, the *Clk* positive feedback loop is likely to be essential for behavioral aspects of circadian rhythmicity. In a second positive feedback loop, *Pdp-1* and *Clk* activate each other (Fig. 1A) (Cyran et al., 2003). Circadian oscillations of *per, vri,* and *Clk* expression are preserved when this loop is removed by fixing *Pdp-1* expression. It is not known whether [PDP-1] oscillations driven by this feedback loop are essential for behavioral rhythmicity.

Yang and Sehgal (2001) generated transgenic *Drosophila* in which the negative feedback loop based on PER's repression of transcription was removed by making *per* and *tim* expression constitutive. In two lines, 40 − 52 % of individuals displayed behavioral rhythmicity and cyclic *vri* expression. The model cannot simulate these observations. With [PER] fixed, the only negative feedback loop is that in which CLK activates *vri* and VRI represses *Clk*. However, this loop does not appear to contain slow kinetic processes able to sustain circadian oscillations. We carried out simulations with this negative feedback loop isolated. Equations 1, 4, 7, 10, 16, 17, and 18 were used, with $[PER_{m, nuc}]$ fixed in Eq. 1. Oscillations could not be obtained, and only steady states were observed. Slow processes such as post-translational protein modifications could be present within this feedback loop, enabling it to sustain a 24-hour oscillation in some individual *Drosophila*. However, it is also plausible that some uncharacterized negative feedback loop drives a "backup" oscillator capable of sustaining a circadian rhythm in some individuals. Normally, this oscillator might be overridden or phase locked by the primary oscillator based on *per* repression.

*Clk* overexpression reduces the simulated period (Fig. 5B). A form of *Clk* overexpression has been described. Kim et al. (2002) expressed *Clk* driven by *per* circadian regulatory sequences in transgenic *Drosophila*. *Per-Clk* mRNA exhibited circadian oscillations ~3-fold higher than, and nearly antiphase (in Zeitgeber time) with endogenous *Clk* mRNA. Nevertheless, the phase of CLK protein oscillations was similar to that in WT flies. The fact that the phase of CLK was not significantly altered despite the large change in the phase of *Clk* mRNA suggests that post-transcriptional regulation of CLK is important in setting the phase of CLK. This result suggests the current model cannot accurately simulate experiments with altered dynamics of *Clk* expression, because CLK post-transcriptional regulation is not represented.

**The model simulates light responses and temperature compensation**

We simulated exposure to light as increasing the degradation rates of cytosolic species of PER. With this assumption, oscillations can be entrained to LD cycles (Fig. 7A). The range of entrainment appears reasonably broad (cycle length of 21 − 25 hrs). A photic phase-response curve (PRC) was also generated (Fig. 7B) which resembles experimental PRCs (Matsumoto et al., 1994; Konopka et al., 1991).

Temperature compensation of oscillation period can be simulated (Fig. 7C) by assuming that a temperature increase decreases the rate of PER nuclear transport. The decrease opposes increased rates of other processes, maintaining a period of ~ 24 hrs. Other authors have suggested this mechanism for temperature compensation. Goldbeter (1995) presented an early model of the *per* negative feedback loop. The simulated period was increased by reducing the rate of PER nuclear transport. This reduction was proposed to contribute to temperature compensation (Hong and Tyson, 1997; Leloup and Goldbeter, 1997). Huang et al. (1995) found that the *perL* mutation, which abolishes temperature compensation, alters the temperature sensitivity of PER dimerization. This altered sensitivity was proposed to change the temperature dependence of the rate of PER nuclear transport, eliminating temperature compensation (Hong and Tyson, 1997).

**The model allows experimental predictions**

Equations 7-15 represent the *per, Pdp-1,* and *vri* expression rates as functions of the acetylation of histone proteins. Fifth powers of the degrees of acetylation are used in Eqs. 7-9 to generate steeply sloped relationships of expression rates *vs.* the levels of transcriptional repressor (PER, which promotes deacetylation) and activator (CLK, which promotes acetylation). Such sigmoidal relationships seem essential for simulation of circadian oscillations. We carried out simulations using only first powers in Eqs. 7-9. Oscillations could not be simulated irrespective of the values of model parameters. Previous models also relied on similar sigmoidal relationships, described by Hill functions of [PER] and [CLK] with Hill coefficients >2 (Ueda et al., 2001; Gonze et al., 2000; Leloup and Goldbeter, 1998; Goldbeter, 1995).

If such relationships are qualitatively correct, experimental predictions follow. Expression of a reporter gene with a *per* or *vri* promoter construct including CLK-CYC binding sites is predicted to depend supralinearly on the amount of CLK-CYC bound to promoters and not complexed with PER. If such a reporter gene could be expressed in an *in vitro* system where [CLK] and [PER] can be varied independently, then for fixed [CLK], the expression rate is predicted to decrease steeply as [PER] is increased (*i.e.*, fitting the curve of expression rate *vs.* [PER] should require powers of [PER] >1). With increasing [CLK], a steep increase of expression might be observed.

Figure 6A illustrates that circadian oscillations of PER persist if positive feedback is eliminated by fixing *Clk* expression. A transgenic *Drosophila* line, based on *Clk*-null mutant animals, might be constructed in which *Clk* expression in lateral neurons is driven by a constitutive promoter. The model predicts that circadian oscillations of PER should still be evident in these neurons, as long as [CLK] is similar to its average during wild-type oscillations. The model also predicts that circadian oscillations of PER, VRI, and CLK could be observed with constitutive *Pdp-1* expression, if [PDP-1] is similar to its average during wild-type oscillations.

Simulation of temperature compensation (Fig. 7C) predicts that an increase in temperature decreases the rate of PER nuclear entry. This prediction might be tested in cultured cells transfected with inducible *per* and *tim* transgenes. The *per* transgene might express a fusion of PER with a reporter such as green fluorescent protein. For different temperatures, the timing and rate of PER nuclear entry could be determined following induction of *per* and *tim*.

Most data to which simulated oscillations are compared are from *Drosophila* whole-head extracts (Bae et al., 2000; Lee et al., 1998). Data is sparse concerning molecular oscillations within *Drosophila* lateral neurons, thought to be the pacemaker cells for the central circadian oscillator (Helfrich-Forster, 1998; Ewer et al., 1992). Oscillations in PER and TIM levels and nuclear localization have been examined in these cells (Shafer et al., 2002), but data concerning oscillations in CLK or other core gene products have not been reported. Such data could necessitate revisions of models describing the *Drosophila* central oscillator, because there is evidence of significant mechanistic differences between central and peripheral oscillators (reviewed in Glossop and Hardin, 2002).

Positive or negative feedback loops, involving interactions between transcriptional activators and repressors, appear to be essential for generation of circadian rhythms in other organisms. In the cyanobacterium *Synechococcus*, a negative feedback loop appears to involve repression of the *kaiC* gene by its product (Iwasaki et al., 2002; Iwasaki and Dunlap, 2000). In mammals, interacting positive and negative feedback loops involve interactions of CLK with isoforms of PER and with cryptochrome proteins (Albrecht, 2002; Shearman et al., 2000). Therefore, models similar to ours should be useful for describing circadian rhythm generation in mammals and other organisms.

## Table I: Standard Parameter Values of the *Drosophila* Model *

| Parameters and Values | Biochemical Significance |
|---|---|
| | |
| $v_{Per} = \lambda * 10.0$ nM hr$^{-1}$, $v_{Vri} = \lambda * 72.0$ nM hr$^{-1}$, $v_{Pdp} = \lambda * 324.0$ nM hr$^{-1}$, $v_{Clk} = \lambda * 1.0$ nM hr$^{-1}$, $R_{Pbas} = \lambda * 0.02$ nM hr$^{-1}$, $R_{Vbas} = \lambda * 0.18$ nM hr$^{-1}$, $R_{Cbas} = \lambda * 0.001$ nM hr$^{-1}$, $R_{PDbas} = \lambda * 0.36$ nM hr$^{-1}$ | Maximal and basal velocities for *per, vri Pdp-1,* and *Clk* expression (Eqs. 13-16). |
| $K_{PV} = 0.2$ nM, $K_{PP} = 0.24$ nM, $K_{PPD} = 0.1$ nM, $K_{VC} = 0.54$ nM, $K_{PDC} = 0.54$ nM, $K_{CV} = 0.083$ nM, $K_{CP} = 0.134$ nM, $K_{CPD} = 0.248$ nM | Binding constants for PER, VRI, PDP-1, and CLK to regulatory elements upstream or within the *per, vri, Pdp-1,* and *Clk* genes (Eqs. 1-3, 16). |
| $v_{pcyt} = \lambda * 1.6$ nM hr$^{-1}$, $K_{pcyt} = 0.25$ nM, $v_{pnuc} = \lambda * 0.3$ nM hr$^{-1}$, $K_{pnuc} = 0.001$ nM, $v_{trans} = \lambda * 1.6$ nM hr$^{-1}$, $K_{trans} = 0.25$ nM, $v_{dclk} = \lambda * 0.2$ hr$^{-1}$, $v_{dvri} = \lambda * 0.7$ hr$^{-1}$, $v_{dPdp} = \lambda * 0.65$ hr$^{-1}$ | Velocities and Michaelis constants for PER phosphorylation, for PER nuclear transport, and degradation of VRI, CLK, and PDP-1 (Eqs. 17-25). |
| $F_V = F_P = F_{PD} = \lambda * 1.0$ hr$^{-1}$, $N = 5$, $k_{Vdeac} = k_{Pdeac} = k_{PDdeac} = \lambda * 0.2$ hr$^{-1}$, $\tau_{Pdp, op} = \tau_{Per, op} = \tau_{Vri, op} = 3.0$ hr $/ \lambda$ | Scaling factors, rate constants, and time constants associated with acetylation and deacetylation of histones (Eqs. 1-6, 10-12). |
| $v_{degp} = \lambda * 5.0$ nM hr$^{-1}$, $K_{degp} = 0.01$ nM | Maximal velocities and Michaelis constants for nuclear degradation of fully phosphorylated PER (Eq. 25). |
| $k_d = \lambda * 0.005$ hr$^{-1}$, $k_{light} = 0.685$ hr$^{-1}$, $\tau_{delay} = 3.0$ hr $/ \lambda$ | Degradation rate constants for all molecular species ($k_d$, Eqs. 17-25) and for PER in response to light ($k_{light}$). Time delay for PDP-1 synthesis ($\tau_{delay}$, Eq. 15). |

* To adjust the free-running oscillation period (Fig. 3) to exactly 24.0 hrs, many parameter values need to be multiplied or divided by a common scaling factor, λ, as shown. The value of λ is 1.062.

**Figure Legends**

*Figure 1*. Schematic of the model for *Drosophila* circadian rhythm generation. A, feedback loops of transcriptional regulation. PER interacts with CLK, inhibiting CLK's activation of *per* and reducing the level of PER, forming a negative feedback loop. *Vri* is activated by CLK, and VRI in turn represses *Clk*. A second negative feedback loop is thus formed. A positive feedback loop also exists in which *Pdp-1* is activated by CLK, and PDP-1 in turn activates *Clk*. A second positive feedback loop is formed from CLK, PER, and VRI as follows: Increased PER binds with CLK and suppresses CLK's activation of *vri*. The level of VRI then falls and *Clk* is thereby de-repressed. Then, more CLK is synthesized and the expression of *per* is further increased. B, the role of PER in more detail. PER undergoes two cytosolic phosphorylations and then enters the nucleus. PER then interacts with CLK, suppressing CLK's activation of *per*. Nuclear PER undergoes further phosphorylations prior to degradation. All phosphorylation states of nuclear PER are assumed competent to interact with CLK (as illustrated by dashed box).

*Figure 2*. Steady-state *per* expression rates implied by the model's description of histone acetylation kinetics and of the effect of acetylation on transcription (Eqs. 1-15). Equations 2, 5, 8, and 13 were used to determine the expression rate as a function of [CLK] and [$PER_{nuc}$], with model parameters at standard values (Table I). A, log-log plot of *per* expression rate *vs.* [CLK] in the absence of repression by PER ([$PER_{nuc}$] = 0). B, log-log plot of *per* expression rate *vs.* [$PER_{nuc}$] with [CLK] fixed at a saturating level (2.5 nM).

*Figure 3*. Simulation of circadian oscillations in constant darkness (DD). Parameters are set to standard values (Table I). A, time courses of [VRI], [$PER_{tot}$], [PDP-1], and [CLK]. B, time courses of PER cytosolic and nuclear species concentrations [$P0_{cyt}$], [$P2_{cyt}$], and [$PER_{nuc}$]. For clarity, time courses of [$P0_{cyt}$] and [$P2_{cyt}$] are scaled vertically by a factor of 10. C, oscillations in [VRI], [$PER_{tot}$], [PDP-1], and [CLK] when $3^{rd}$ powers of $AC_{Vri}$, $AC_{Per}$, and $AC_{Pdp}$ are used in Eqs. 7-9. Values of [CLK] are multiplied by 2 for ease of visualization. D, periods and amplitudes of simulated circadian oscillations in [$PER_{tot}$]. To generate these oscillations, each parameter in the set of standard parameter values (Table I, excluding N and $k_{light}$) was increased or decreased by 20%. There are 77 data points including the control with all parameter values standard. The control is at the intersection of the horizontal and vertical red lines.

*Figure 4*. Simulations of clock gene mutants. Parameters are at standard values (Table I) except as specified for each simulation. In Fig. 4 and subsequently, time courses of [CLK], [VRI], [$PER_{tot}$], and [PDP-1] are respectively solid black, dashed black, solid grey, and dashed grey. A, a *per[01]* homozygous mutation. PER is assumed unable to interact with CLK or regulate transcription. This is implemented by setting the Hill functions of [$PER_{nuc}$] to 1 in Eqs. 1-3. B, a *Clk[Jrk]* homozygous mutation. CLK is assumed unable to activate transcription. This is implemented by setting the Hill functions of [CLK] to zero in Eqs. 1-3. C, decrease in period with increased PER phosphorylation. The maximal phosphorylation velocities and the nuclear transfer velocity ($v_{pcyt}$, $v_{pnuc}$, and $v_{trans}$) were increased by 65% from their standard values.

*Figure 5.* Simulations with altered *vri* and *Clk* expression. A, a *vri^null*/+ mutant. The VRI synthesis rate $R_{vri}$ (Eq. 14), was scaled by a factor of 0.5 throughout the simulation. All species continue to show large-amplitude oscillations. The period is 23.3 hrs. B, *Clk* overexpression. A constant value of 0.3 nM hr$^{-1}$ was added to the CLK synthesis rate $R_{clk}$ (Eq. 16). The oscillation period is 23.4 hrs.

*Figure 6.* Simulations with fixed *Clk* and *vri* expression. All parameters are at standard values except as specified. A, fixed *Clk* expression. [CLK] is fixed at half of its maximal concentration in Fig. 3A (at 0.79 nM). Oscillations in [PER$_{tot}$], [VRI], and [PDP-1] persist, with period near circadian (26.0 hrs). B, constitutive *vri* expression is simulated by fixing [VRI] at 3.0 nM. [PER$_{tot}$] and [PDP-1] continue to oscillate, and oscillations in [PDP-1] drive low-amplitude oscillations in [CLK]. The plotted [CLK] values are increased 10-fold for visualization.

*Figure 7.* Simulated effects of light exposure and simulated temperature compensation. All parameters are at standard values except as specified. A, entrainment of oscillations to a 22-hr light-dark cycle. The overbar indicates the dark phase. Light is assumed to increase the degradation of cytosolic PER. During the light phase, a first-order degradation term was added to the right-hand side of the differential equations for [P0$_{cyt}$], [P1$_{cyt}$], and [P2$_{cyt}$]. The first-order rate constant was 0.685 hr$^{-1}$. B, photic phase-response curves (PRCs). When constructing the model PRC (solid curve), each light pulse was simulated by adding, for 2.0 hrs, a first-order degradation term to the differential equations for each cytosolic form of PER. The first-order rate constant was 0.685 hr$^{-1}$. An experimental *Drosophila* PRC (Fig. 5 of Konopka et al., 1991) is also illustrated (circles). The means of the experimental phase shifts are displayed. C, temperature-compensated oscillations. The standard parameter value set (Table I) was used as the starting point. Parameters containing units of time (hr) or inverse time (hr$^{-1}$) were respectively divided or multiplied by a $Q_{10}$ of 2.0 to simulate a temperature increase of 10° C. However, the maximal velocity for PER nuclear transport ($v_{trans}$) was assumed to decrease with increasing temperature, and was multiplied by a $Q_{10}$ of 0.62. A period of 24.2 hrs results. Values of [CLK] are multiplied by a factor of 3 for ease of visualization.

A

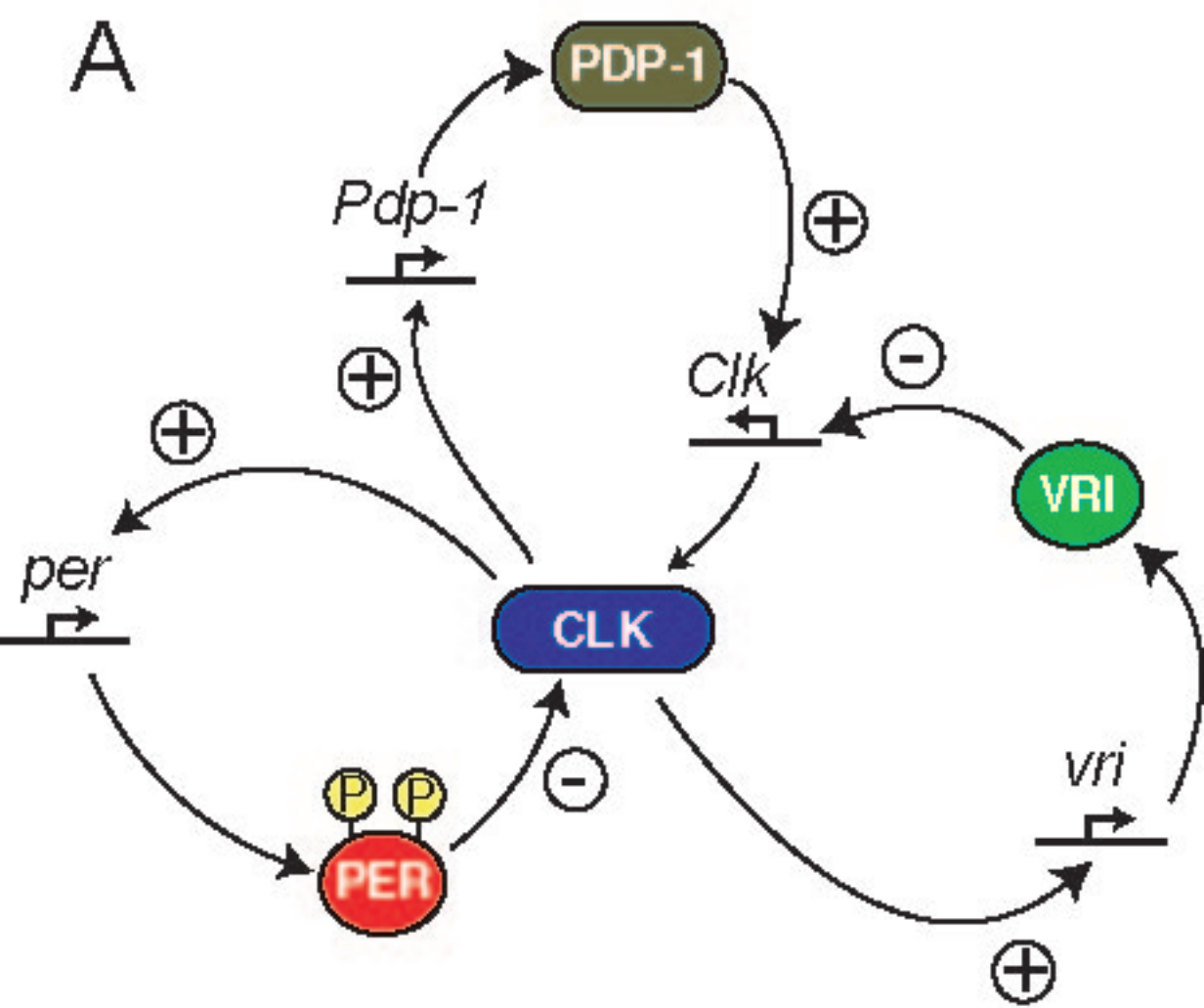

B

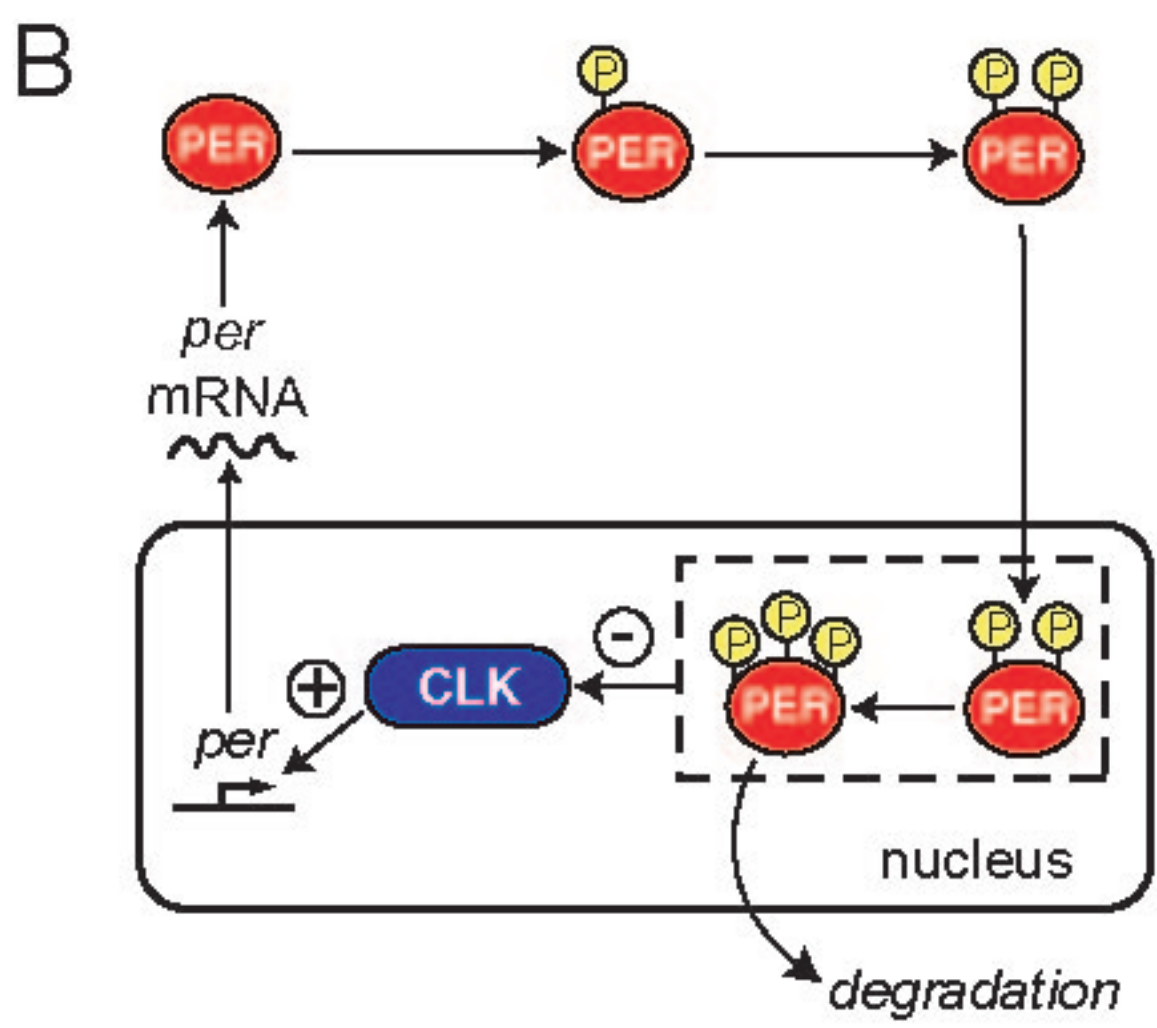

# Figure 2
# P. Smolen et al.

A

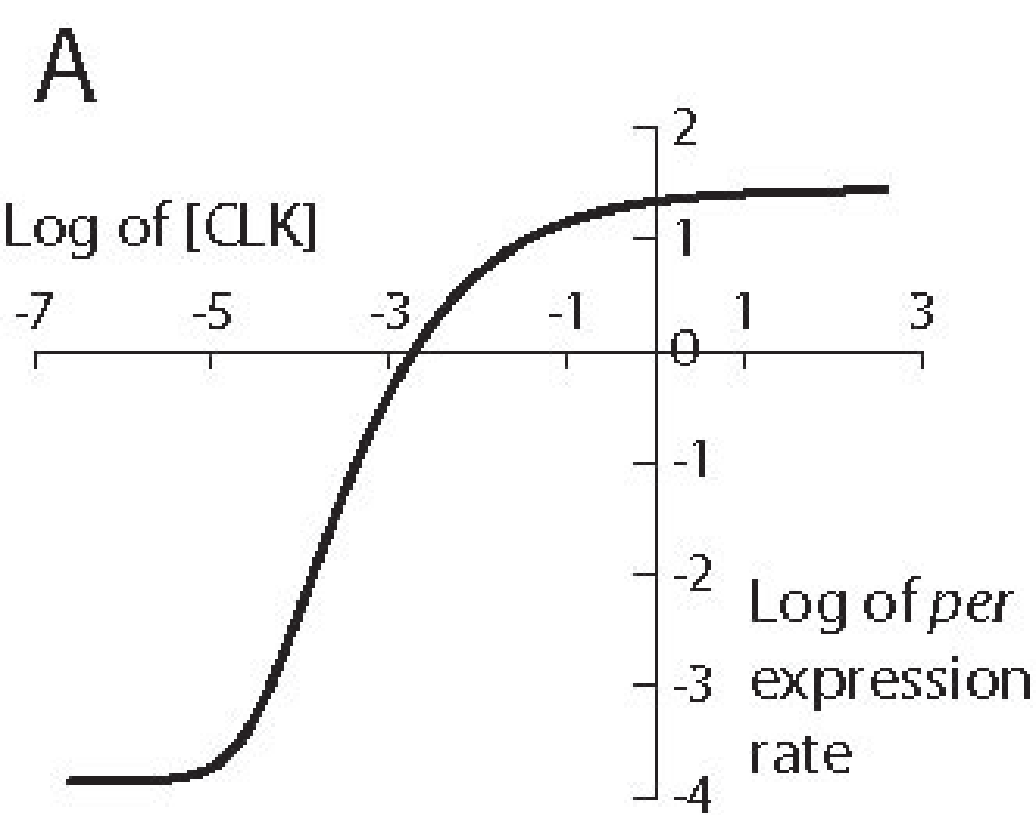

Log of [CLK]

Log of *per* expression rate

B

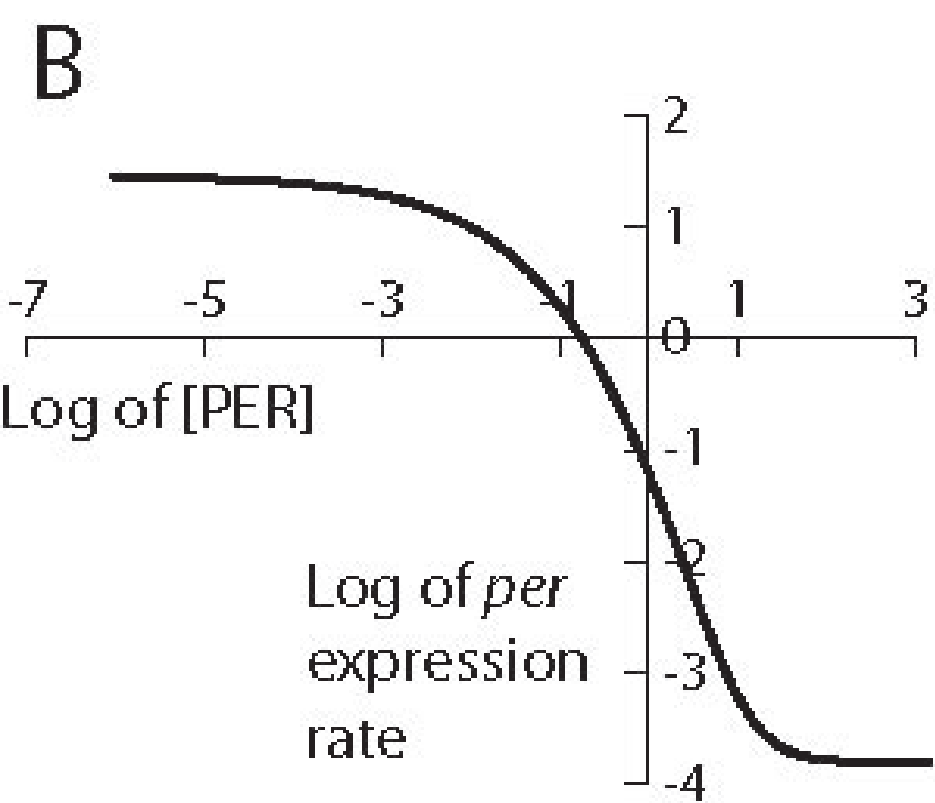

Log of [PER]

Log of *per* expression rate

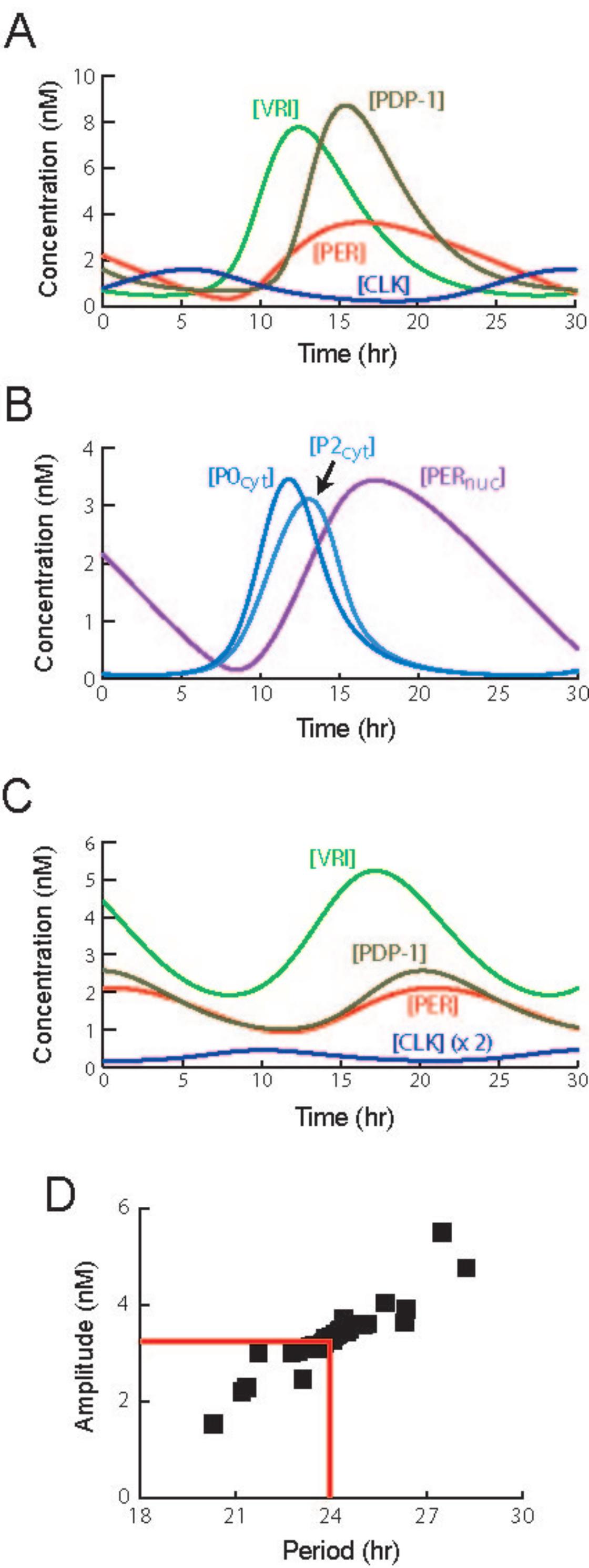

## A. *per⁰¹*

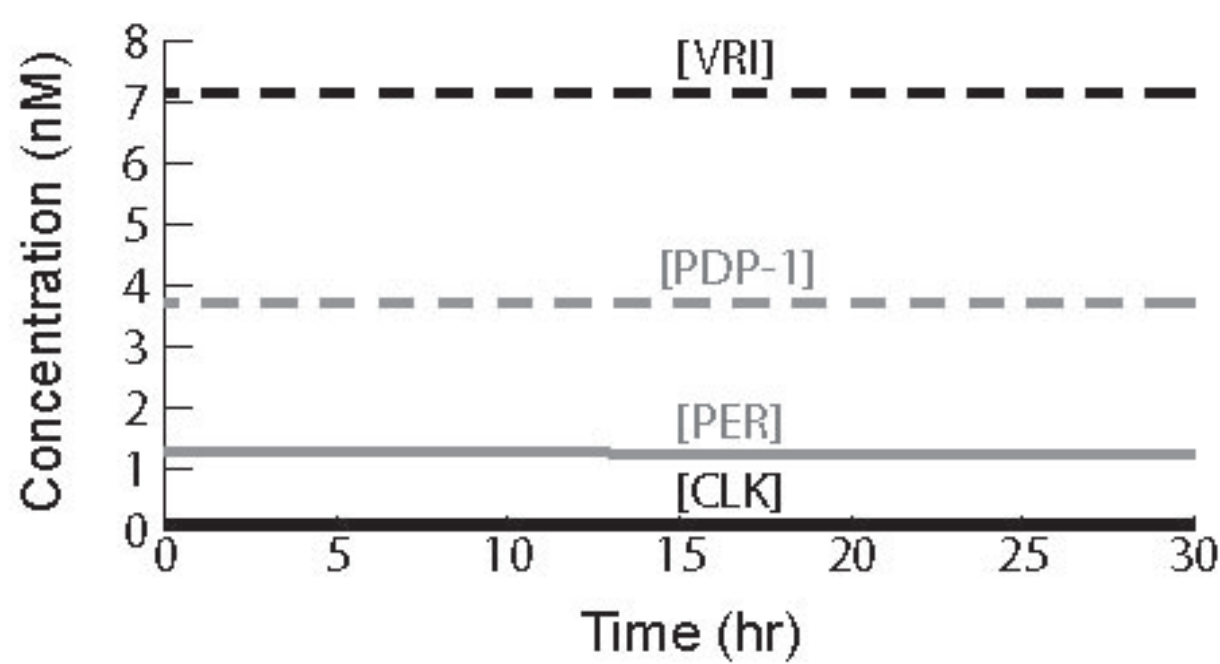

## B. *Clk^Jrk*

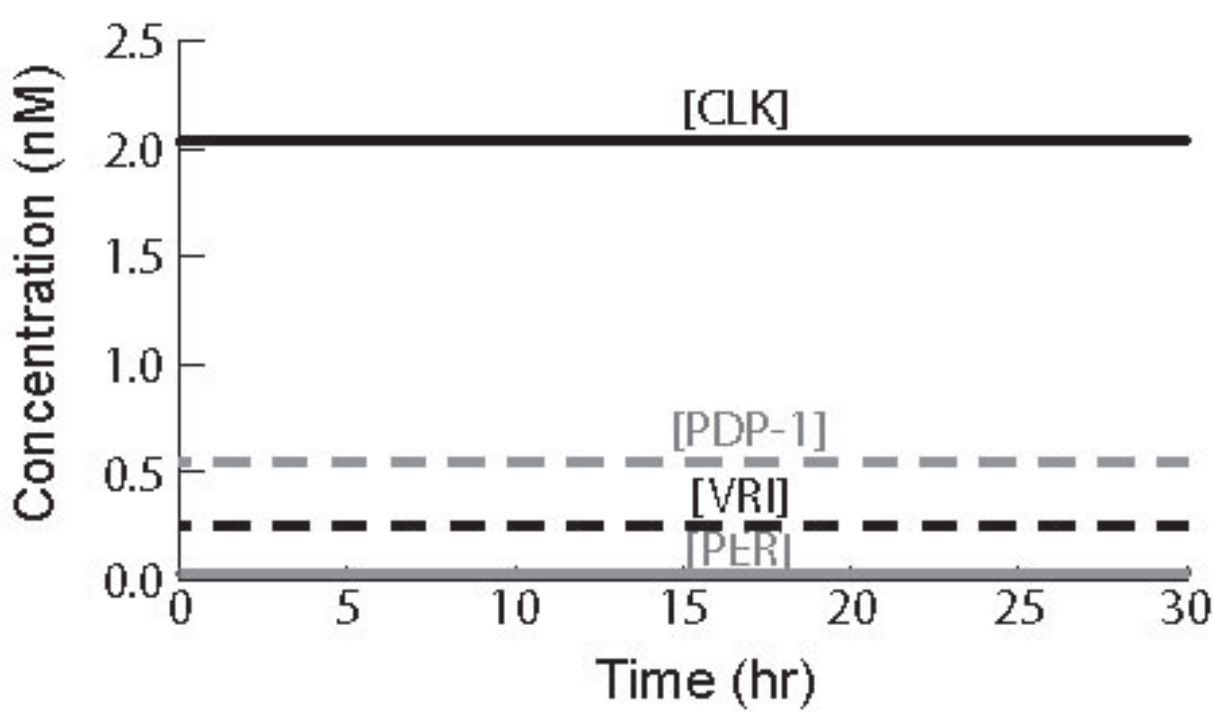

## C. Increased PER phosphorylation

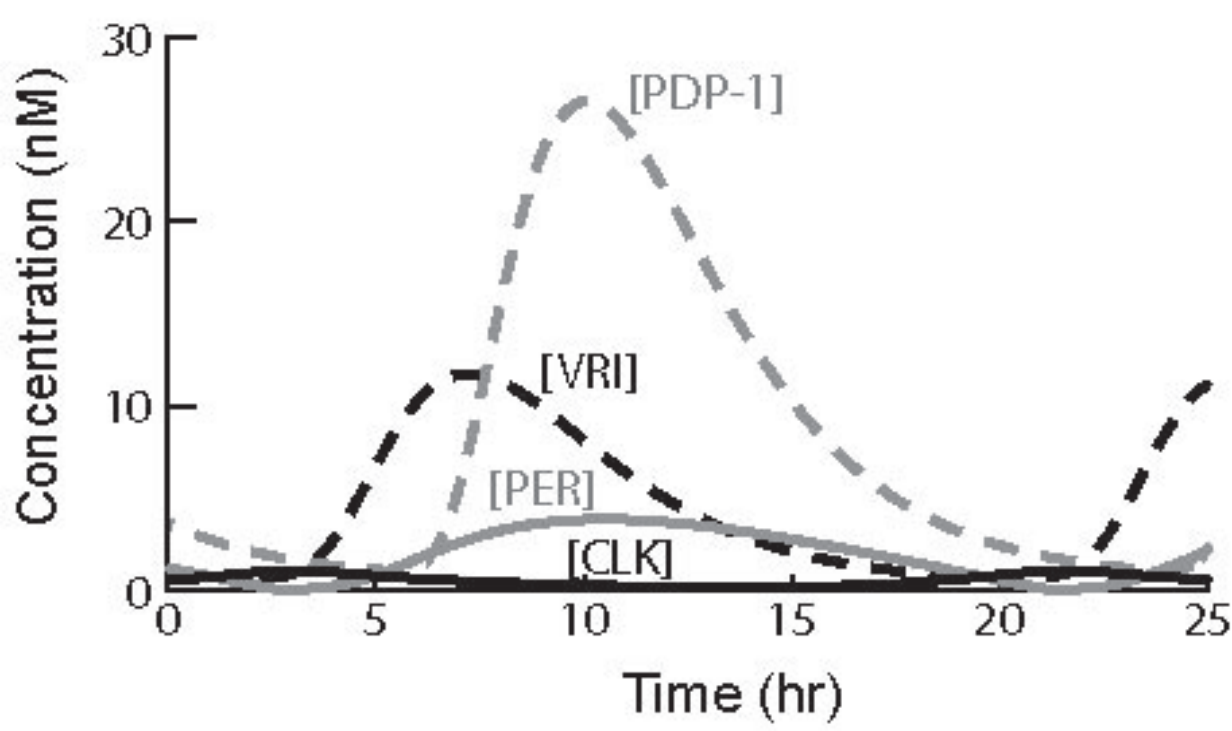

## A. *vri*<sup>null</sup>/+

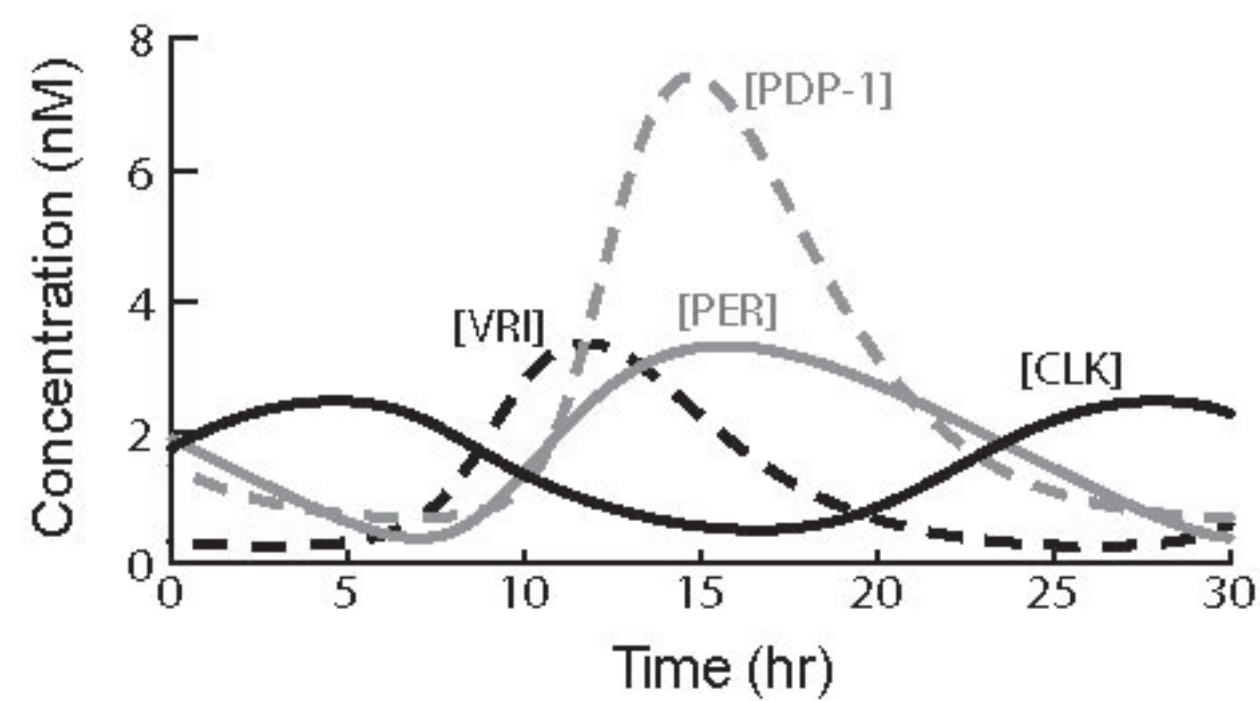

## B. CLK overexpression

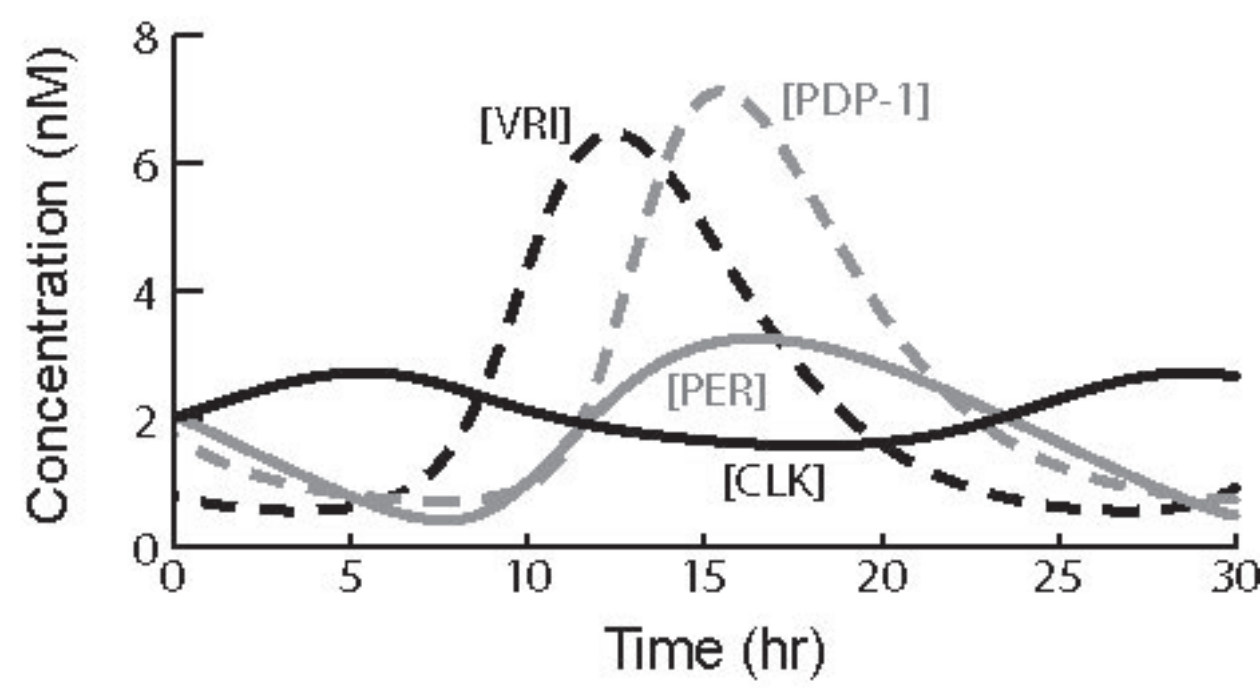

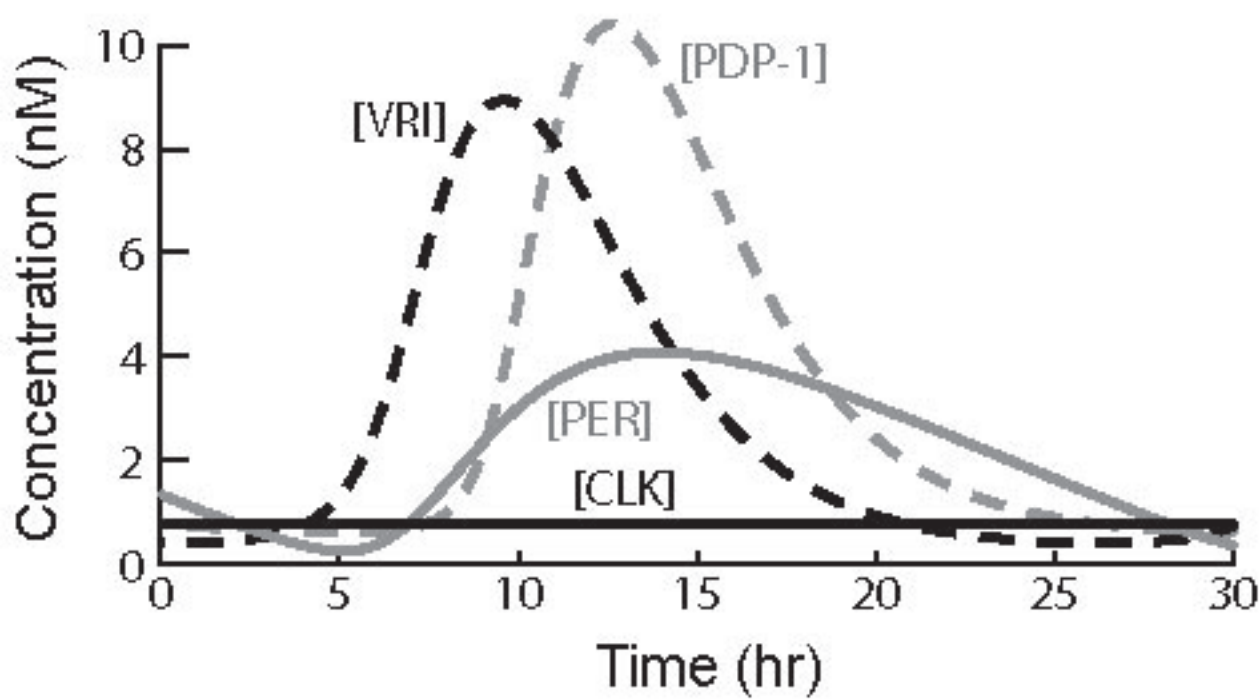

## A. Constitutive *Clk*

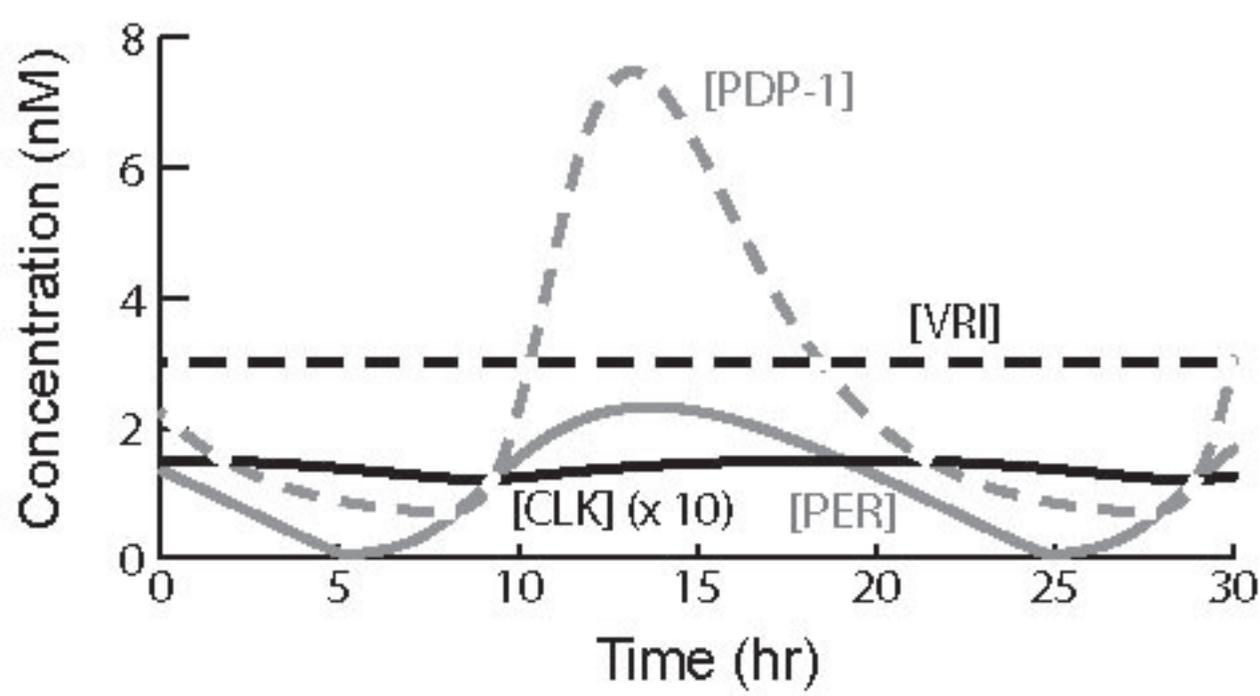

## B. Constitutive *vri*

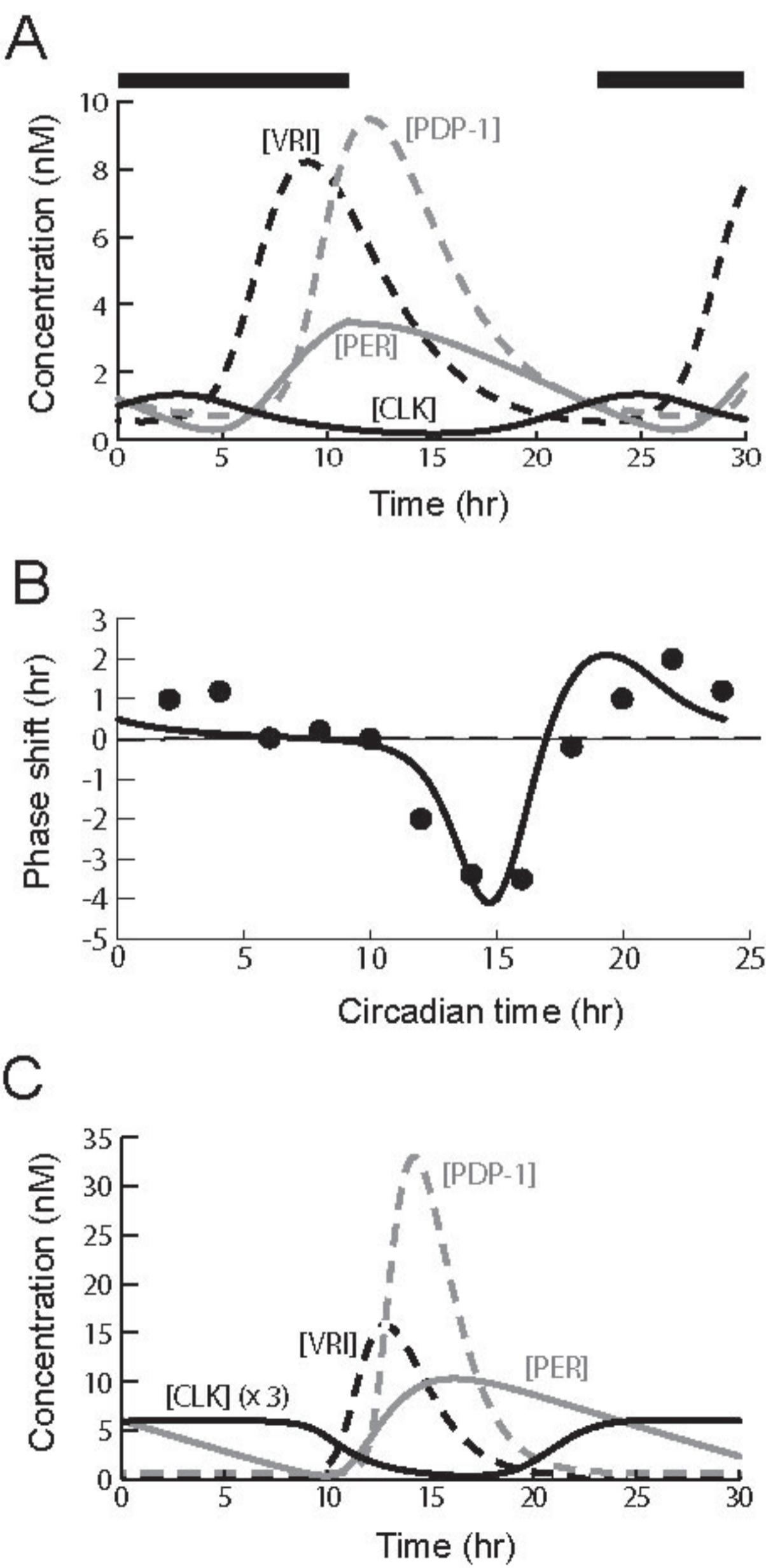